\documentclass[aps,prx,onecolumn,amsmath,superscriptaddress,nofootinbib,amssymb,11pt]{revtex4-1}
\usepackage{graphicx}
\usepackage[colorlinks, linkcolor=blue]{hyperref}
\usepackage{verbatim}
\usepackage{latexsym}
\usepackage{amsmath}
\usepackage{amssymb}
\usepackage{hyperref}
\usepackage[font=small,labelfont=bf]{caption}
\usepackage{subcaption}

\def \v {{\bf v}}

\def \Q{{\vec Q}}

\def \beq {\begin{eqnarray}}
\def \eeq {\end{eqnarray}}
\def \tn {\textnormal}

\renewcommand{\vec}[1]{\boldsymbol{#1}}

\newcommand{\ben}{\begin{equation}}
\newcommand{\een}{\end{equation}}
\newcommand{\ba}{\begin{array}{ccc}}
\newcommand{\ea}{\end{array}}
\newcommand{\nn}{\nonumber \\}

\begin{document}

\title{Confinement transition to density wave order in metallic doped spin liquids}

\author{Aavishkar A. Patel}
\author{Debanjan Chowdhury}
\author{Andrea Allais}
\affiliation{Department of Physics, Harvard University, Cambridge Massachusetts
02138, USA.}
\author{Subir Sachdev}
\affiliation{Department of Physics, Harvard University, Cambridge Massachusetts
02138, USA.}
\affiliation{Perimeter Institute of Theoretical Physics, Waterloo Ontario-N2L 2Y5, Canada.}

\date{\today \\
\vspace{1.6in}}
\begin{abstract}
Insulating quantum spin liquids can undergo a confinement transition to a valence bond solid via the condensation of topological excitations of the associated gauge theory. We extend the theory of such transitions to fractionalized Fermi liquids (FL*): these are
metallic doped spin liquids in which the Fermi surfaces only have gauge neutral quasiparticles. Using insights from a duality transform on a
doped quantum dimer model for the U(1)-FL* state, we show that projective symmetry group of the theory of the topological excitations remains unmodified, but the Fermi surfaces can lead to additional frustrating interactions. We propose a theory for the confinement transition of $\mathbb{Z}_2$-FL* 
states via the condensation of visons. A variety of confining, incommensurate density wave states are possible, including some
that are similar to the  incommensurate $d$-form factor density wave order observed in several recent experiments on the cuprate superconductors. 
\end{abstract}

\maketitle

\section{Introduction}
The cuprate superconductors at low doping display a number of complex phenomena \cite{LNW06}. Below the ``pseudogap" temperature ($T^*$), the metallic state displays Fermi-liquid like behavior \cite{Marel13, MG14} but is unlike any conventional metal in that the carrier density is inconsistent with the total Luttinger count \cite{PJ11}. On the other hand, the Fermi-liquid state seen at large values of the doping has been studied extensively \cite{MHJCP13} and satisfies Luttinger's theorem. A description of the transition between the distinct metallic states and its relation to the phenomenology of the ``strange-metal" continues to be elusive. Much of the recent activity in the field has been devoted to a study of the ubiquitous charge-density wave (CDW), observed in a number of different families of the underdoped cuprates \cite{Ghiringhelli12,DGH12,Chang12,comin13,neto13,DGH13,MHJ11,MHJ13,DLB13, JH02,AY04,lawler,mesaros,Comin14sym,Fujita14}. The incommensurate charge-density wave state onsets at a temperature below $T^*$, but above the superconducting $T_c$. The relationship, if any, between the pseudogap metal and the CDW is a topic of great interest \cite{DCSSrev}.

In many of the spin-liquid based approaches, there is a parent state that describes the metallic state in the absence of any broken symmetries \cite{LNW06,RK07,RK08,SS09,YQSS10,PAL14,PAS15}, and represents a deconfined phase of an appropriately defined gauge theory. The ordered phases observed at lower temperatures are then interpreted as instabilities arising out of this state. Two of the present authors studied the `weak-coupling' instabilities of a particular candidate state---fractionalized Fermi-liquid (FL*)--- in the presence of short-range interactions and reproduced many of the experimentally observed trends associated with the CDW \cite{DCSS14b}.  However, the resulting metallic CDW state in this earlier study is more properly referred 
to as CDW*, in the notation of Ref.~\onlinecite{TSMPA00}: this is because 
the deconfined gauge excitations of the FL* state remain largely unmodified
across the transition to charge order. 

The present paper will take an alternative view of the onset of charge order in the FL* state: we will present a theory in which
the appearance of charge order {\it co-incides\/} with a confinement transition in the gauge theory; so our confining phase
will be a true CDW and not a CDW*. Such co-incident transitions
have been well studied in early work on insulating spin liquids \cite{rsprl1,rsprb1,RJSS91,CSS93,SSMV99,TSMPA00,senthil1,senthil2}.
Here, we will extend such theories to metallic states, and show that the FL* to confining-CDW transition has the same general
structure as the corresponding transition in the insulator. At first sight, this similarity should appear surprising. For the case of conventional Landau-Ginzburg transitions, it is well-known that the theory for the onset of broken symmetry in insulators
is very different from that in a metal: the presence of the Fermi surface over-damps the order parameter fluctuations, and this changes
the nature of the critical fluctuations \cite{Hertz76}. However, this feature does not extend to confining transitions in gauge theories
because there is no `Yukawa' coupling between the order parameter and the gapless Fermi surface excitations \cite{Morinari02,Kaul08,TGTS09}. Furthermore, we will show here that the Berry phase terms in the gauge theory (which are responsible 
for the charge order at the confinement transition) retain the same form in the FL* metal as that in the insulator \cite{Kaul08};
this is due to the absence of any gauge-charged quasiparticles on the Fermi surface.
Consequently the projective symmetry transformations constraining the effective theory for the topological excitations
also have the same form in the insulator and the FL*. We will argue, therefore, that the primary effect of the presence
of the Fermi surface is that it can generate longer-range and frustrating couplings in the action for the topological excitations.
These longer-range couplings can, in turn, lead to a richer set of possibilities \cite{CXLB11} for the structure of the charge ordering
in the confining phase. 

Recent low-temperature and high-field measurements of the Hall-coefficient \cite{LTCP15} have accessed the 
metallic ground state in the absence of superconductivity near optimal doping. 
This study has reported some interesting, and perhaps, surprising results. As a function of decreasing hole-doping ($x$), the experiments are consistent with two separate transitions. At higher-doping ($x_h\approx19\%$), there is a transition from a conventional Fermi-liquid, with $1+x$ carriers, to a metallic state with $x$ carriers and no broken translational symmetry. The metallic state with only $x$ carriers would be consistent with an FL* with reconstructed hole-pockets; the presence of a background spin-liquid is crucial for the reconstruction and violation of Luttinger's theorem \cite{MO00,TSMVSS04}. The subsequent transition at lower doping ($x_l\approx16\%$), corresponds to the onset of quasi long-ranged charge density wave. 

In order to make sharp theoretical statements, we shall focus on the metallic ground states at $T=0$. Assuming the intermediate metallic state ($x_l<x<x_h$) is described by a FL*, it is natural to ask if the onset of broken translational symmetry at $x=x_l$ could be concomitant with a confinement transition. In this paper, we shall primarily focus on the case of a $\mathbb{Z}_2$-FL* characterized by background topological order, which can, in principle, survive as a stable ground state in (2+1)-dimensions at $T=0$. 
We will be interested specifically in studying the effect of condensing the excitations carrying the $\mathbb{Z}_2$ magnetic flux \cite{RJSS91}, the visons, and the associated patterns of broken translational symmetry that this generates. 
 
The rest of this paper is organized as follows. We begin in Section~\ref{sec:u1} by describing an explicit duality transformation on a model of metallic
doped spin liquid: the U(1)-FL* state described by 
the quantum dimer model of Ref.~\cite{PAS15}. The general lessons from this analysis will be employed in the subsequent sections
for a detailed study of a dual theory of a $\mathbb{Z}_2$-FL* state: this theory will be described in Section~\ref{sec:z2}, and its phase
diagram will be presented in Section~\ref{sec:results}.

\section{U(1) FL* from a quantum dimer model}
\label{sec:u1}

We begin by deriving the effective field theory of the U(1)-FL*, obtained in the dimer model construction of Ref.~\cite{PAS15}. 
This model can be extended to obtain a $\mathbb{Z}_2$-FL* by allowing for non-nearest-neighbor dimers, as in the insulator \cite{SSkagome,SSMV99,MS01}, and we will discuss this further in Section~\ref{sec:z2}.

\begin{figure*}[!ht]
\begin{subfigure}[b]{3.1in}
		\includegraphics[width=3.0in]{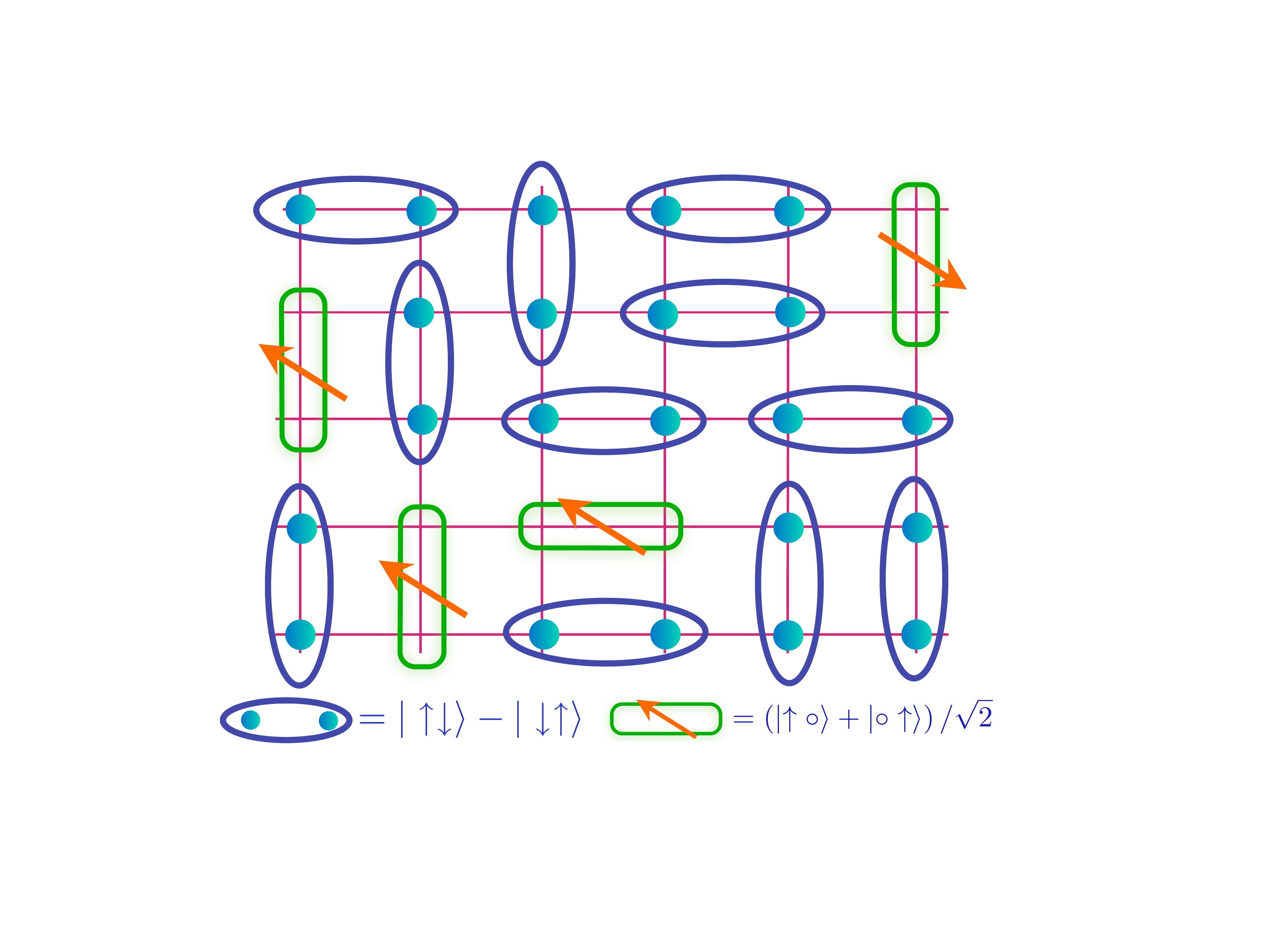}
		\caption{}
	\end{subfigure}
\begin{subfigure}[b]{3.1in}
		\includegraphics[width=3.0in]{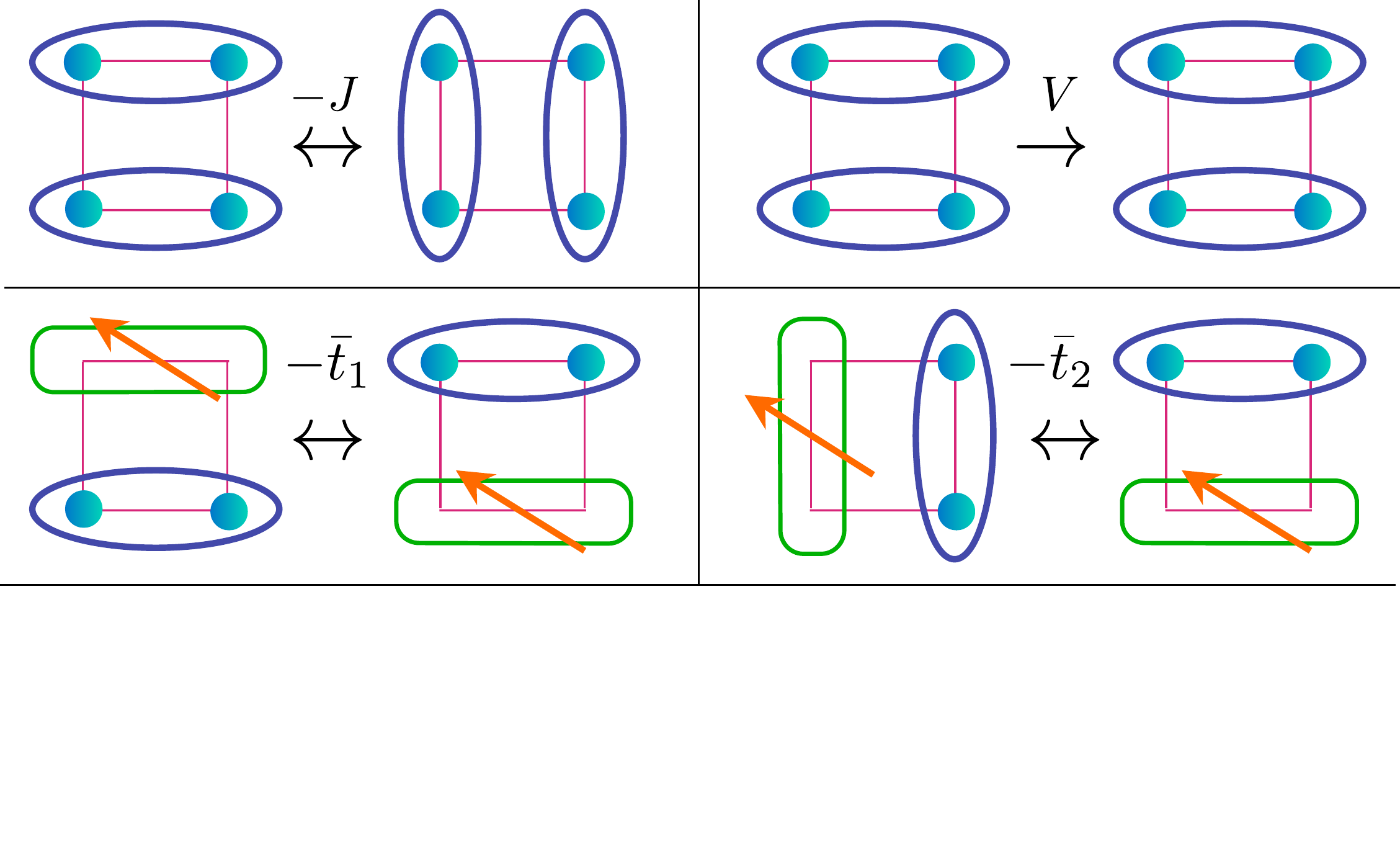}
		\caption{}
	\end{subfigure}
\caption{(a) A snapshot of the FL* configuration \cite{PAS15}, where the bosonic (fermionic) dimers are shown in blue (green). The gauge-neutral fermionic dimers arise as a result of binding between spinons and holons. The density of the fermionic dimers is $x$, while the total hole-concentration is $(1+x)$.  (b) Pictorial illustration of the various terms in the dimer model defined by  Eq.~(\ref{dham}).}
\label{flstar}
\end{figure*}

The dimer model has interacting bosonic and fermionic dimers at close-packing on the square lattice 
(see Fig. \ref{flstar} and Ref. \cite{PAS15}). This describes a U(1)-FL* phase obtained upon doping away from the Rokhsar-Kivelson insulating point \cite{DRSK88}. At $T=0$ and in (2+1)-dimensions, the U(1) FL* does not represent a stable fixed-point, and there is a flow
to a confining state with CDW order, as we will describe below. 
Nevertheless, the purpose of this exercise is to derive a field-theory for such a phase and highlight the modifications that arise in the usual dimer description of the insulating spin-liquid \cite{SSMV99}, when doped with fermionic dimers.

\subsection{Lattice Hamiltonian}

We begin by recalling the dimer model Hamiltonian of Ref.~\cite{PAS15}:
\beq
H_D &=& H_{\rm RK} + H_{D1} \nn
H_{\rm RK} &=& \sum_i \left[ -J \, D_{ix}^\dagger D_{i+\hat{y},x}^\dagger D_{iy}^{\vphantom\dagger} D_{i+\hat{x},y}^{\vphantom\dagger} + \mbox{~1 term} +V \, D_{ix}^\dagger D_{i+\hat{y},x}^\dagger D_{ix}^{\vphantom\dagger} D_{i+\hat{y},x}^{\vphantom\dagger} + \mbox{~1 term} \right] \nn
H_{D1} &=& \sum_i \left[ -\bar{t}_1 \, D_{ix}^\dagger F_{i+\hat{y},x s}^\dagger F_{ix s}^{\vphantom\dagger} D_{i+\hat{y},x}^{\vphantom\dagger} + \mbox{~3 terms}  -\bar{t}_2 \, D_{i+\hat{x},y}^\dagger F_{iys}^\dagger F_{ixs}^{\vphantom\dagger} D_{i+\hat{y},x}^{\vphantom\dagger} + \mbox{~7 terms}  \right], \label{dham}
\eeq
where the undisplayed terms are generated by operations of the square lattice point group on the terms above. 
Here the $D_{i\alpha}$ are the bosonic dimers, $F_{i\alpha s}$ are the fermionic dimers with spin $s=\uparrow, \downarrow$. 
We identify the dimers by site, $i$, of the square lattice on their lower or left end, and the direction $\alpha = x,y$.
The first term, $H_{\rm RK}$, co-incides with the RK model for the undoped dimer model at $x=0$. Single fermion hopping terms are contained in $H_{D1}$, with hoppings $\bar{t}_{1,2}$ which are expected in the mapping from a $t$-$J$ model \cite{PAS15}. These terms are shown pictorially in Fig.~\ref{flstar}(b). We have only retained terms which
operate on a single plaquette, and these are also the terms which can be included in the explicit duality mapping.

As a first step, we rewrite $H_D$ in a form which makes the connection to a compact U(1) gauge theory evident.
We introduce \cite{fradkiv90,FradkinJHU} an integer-valued `electric field' operator $\hat{E}_{i \alpha}$ on each link of the square lattice,
so that $\eta_i \hat{E}_{i\alpha}$
is the number operator for the dimer on site $i$ oriented in the $\alpha$
direction; $\eta_i$ indicates the sublattice of
site $i$, and equals $+1$ on one sublattice and $-1$ on the other.
Note that $\eta_i \hat{E}_{i\alpha}$ counts the number of {\it both\/} fermionic and bosonic dimers, 
\ben
\eta_i \hat{E}_{i\alpha} = D_{i \alpha}^\dagger D_{i \alpha}^{\vphantom\dagger} + F_{i \alpha s}^\dagger F_{i \alpha s}^{\vphantom\dagger},
\een
and so
the constraint that there  will be exactly one dimer emerging from
every site can be written as
\begin{equation}
\Delta_{\alpha} \hat{E}_{i \alpha} = 2S\eta_i ,
\label{f1}
\end{equation}
where $\Delta_{\alpha}$ is the discrete lattice derivative in the $\alpha$
direction, and $2S=1$. We have introduced a general integer $2S$ for generalization to the case of spin-$S$ antiferromagnets.
The factors of $\eta_i$ were introduced so that the constraint
would have the Gauss-law form in Eq.~(\ref{f1}).
We also introduce an angular phase variable, $\hat{A}_{i\alpha}$ (the analog of a compact $U(1)$ gauge field),
on every link which is canonically conjugate to $\hat{E}_{i
\alpha}$:
\begin{equation}
[\hat{A}_{i \alpha}, \hat{E}_{j \beta} ] = i \delta_{ij}
\delta_{\alpha \beta},
\label{f2}
\end{equation}
where it should be clear from the context when we mean $i =
\sqrt{-1}$,
and when $i$ is a site label. The operator $e^{i \hat{A}_{i \alpha}}$ is then a dimer creation operator, and it is related to the bosonic
and fermionic dimer operators by
\beq
D_{i \alpha}^\dagger &=& e^{i \hat{A}_{i \alpha}} \nn
F_{i \alpha s}^\dagger &=& e^{i \hat{A}_{i \alpha}} f_{i \alpha s}. \label{fqed}
\eeq
Here is $f_{i \alpha s}$ is a gauge-neutral fermionic operator which has the same quantum numbers as an unfractionalized electron residing on the link
$i\alpha$; so we will often refer to $f_{i \alpha s}$ simply as an `electron'.
We can now map $H_D$ into the form of a compact U(1) gauge theory \cite{fradkiv90,FradkinJHU}:
\beq
H_{\rm qed} &=& H_d + H_f + H_{df}\\
H_d &=& \frac{K_1}{2} \nonumber\sum_{i,\alpha} \hat{E}_{i \alpha}^2 - K_2 \sum_i \cos(\epsilon_{\alpha\beta}\Delta_{\alpha}\hat{A}_{i \beta}), \\
H_{f} &=& \sum_i \left[ -t_1 \,  f_{i+\hat{y},x s}^\dagger f_{ix s}^{\vphantom\dagger}  + \mbox{~3 terms}  
- t_2 \,  f_{iys}^\dagger f_{ixs}^{\vphantom\dagger}  + \mbox{~7 terms}  \right] \nn
H_{df} &=& \sum_{i, \alpha} \hat{E}_{i \alpha}\, \mathcal{G}_{i \alpha}  + \sum_{i} \left[\exp \left( i\epsilon_{\alpha\beta} \Delta_{\alpha}\hat{A}_{i \beta} \right) \mathcal{K}_a + \textnormal{H.c.}\right].
\label{f3}
\eeq
The dynamics of the bosonic dimers in $H_{\rm RK}$ are described by $H_d$, which is the same as that in Ref.~\cite{SSMV99}:
the first term, proportional to $K_1$ is only non-trivial when $2S > 1$, and it ensures
that the density of dimers is as uniform as possible. It follows
from the commutation relations (\ref{f2}) that the second term,
proportional to $K_2$, flips dimers around a plaquette; this is the same as the action of the $J$ term in $H_{\rm RK}$, and in perturbation theory
the value of $K_2$ is proportional to $J$.
The hopping Hamiltonian for the electrons, $H_f$, has hopping terms which descend directly from the terms in $H_{D1}$.
Finally, $H_{df}$ contains new terms coupling the dimers to the fermions: the $\mathcal{G}_{i \alpha}$ and $\mathcal{K}_a$
represent bi-linears of the $f_{i \alpha s}$ consistent with the symmetries of the underlying lattice. 
Here $a$ is site of the
dual lattice with co-ordinates $a = (a_x, a_y)$ and $a_{x,y}$ integers. In the present case, $a$ resides at the center of
plaquette around which the `flux' $\epsilon_{\alpha\beta} \Delta_{\alpha}
\hat{A}_{i \beta}$ resonates the dimers. We also introduce the vectors $\hat{e}_x = (1/2,0)$ and $\hat{e}_y = (0, 1/2)$.

\subsection{Dualities and Height model}

We will now write down a path integral representation of the
partition function of $H_d+H_{df}$ by following a standard route  \cite{SSMV99}. We
insert complete sets of $\hat{E}_{i \alpha}$ eigenstates at small
imaginary time intervals $\Delta \tau$. The matrix elements of the
`trigonometric' terms in $H_d+H_{df}$ are evaluated by replacing it with the
Villain periodic Gaussian form. For this we manipulate the action by keeping
all terms second order in $\hat{A}$ and $\mathcal{K}$, while respecting the periodicity $\hat{A} \rightarrow \hat{A} + 2 \pi$:
\begin{eqnarray}
&& \exp\Biggl( K_2 \Delta \tau \cos(\epsilon_{\alpha\beta} \Delta_{\alpha}
\hat{A}_{i \beta}) - \Delta \tau \left[\exp \left( i \epsilon_{\alpha\beta} \Delta_{\alpha}
\hat{A}_{i \beta} \right) \mathcal{K}_a + \mbox{H.c.}\right]\Biggr) \nonumber \\
&&~~\approx \exp\Biggl( K_2 \Delta \tau \cos \left(\epsilon_{\alpha\beta} \Delta_{\alpha}
\hat{A}_{i \beta} + i \frac{( \mathcal{K}_a - \mathcal{K}_a^\dagger)}{K_2} \right)
- \Delta \tau \frac{( \mathcal{K}_a - \mathcal{K}_a^\dagger)^2}{2 K_2} 
 \Biggr) \nonumber \\
&&~~\approx \sum_{p_a} \exp\Biggl( - \frac{K_2 \Delta \tau}{2} \left(\epsilon_{\alpha\beta} \Delta_{\alpha}
\hat{A}_{i \beta} + i \frac{( \mathcal{K}_a - \mathcal{K}_a^\dagger)}{K_2} - 2 \pi p_a \right)^2
- \Delta \tau \frac{( \mathcal{K}_a - \mathcal{K}_a^\dagger)^2}{2 K_2} 
 \Biggr) \nonumber \\
&&~~ = \sum_{B_a} \exp \left(
- \frac{\left(B_a  + \Delta \tau (\mathcal{K}_a - \mathcal{K}_a^\dagger) \right)^2}{2 K_2 \Delta \tau} + i B_a \epsilon_{\alpha\beta} \Delta_{\alpha}
\hat{A}_{i \beta} \right),
\label{f4}
\end{eqnarray}
where $p_a$ and $B_a$ are integer-valued fields on the dual lattice sites, $a$.

A three-vector notation in space time will also be useful: we define the
integer-valued `electromagnetic flux' vector
\beq
F_{a\mu} = (E_{iy}, -E_{ix}, -B_a) 
\eeq
on the dual lattice sites, where
the index $\mu = (x,y,\tau)$ (we will consistently use the labels
$\alpha,\beta \ldots$
to represent spatial components only, while $\mu,\nu,\lambda\ldots$ will represent
three-dimensional spacetime components). Here $E_{i \alpha}$ refer to the
integer eigenvalues of the operator $\hat{E}_{i \alpha}$ which are summed
over in each time step. After performing the integral over the $\hat{A}_{i \alpha}$
we obtain the partition function (we drop the fermion kinetic energy terms in $H_f$ below)
\begin{eqnarray}
Z_1 = \sum_{\{F_{a \mu}\}} && \exp \left( - \sum_a\frac{\left(F_{a\tau}  - \Delta \tau (\mathcal{K}_a - \mathcal{K}_a^\dagger) \right)^2}{2 K_2 \Delta \tau} - \sum_{a, \alpha}\Delta \tau \left[ \frac{K_1 }{2}  F_{a \alpha}^2 + F_{a \alpha} \mathcal{G}_{a \alpha}\right] \right) \nonumber \\
&&~~~~~~~~~~~~~~~~\times \prod_{a,\mu} \delta \left(\epsilon_{\mu\nu\lambda}\Delta_{\nu} F_{a \lambda} - 2S \eta_i \delta_{\mu\tau} \right),
\label{f5}
\end{eqnarray}
where
\beq
\mathcal{G}_{a \alpha} = \epsilon_{\alpha\beta} \mathcal{G}_{i \beta}.
\eeq
The sum in $Z_1$ is over the integer-valued field $F_{a \mu}$ which
resides on the sites of the dual cubic lattice in spacetime; the
delta function constraint imposes `Gauss's law' (Eq.~(\ref{f1})).

By carrying out standard manipulations, as summarized in Appendix \ref{app:duality},
 we obtain a sine-Gordon theory coupled to the fermions
\beq
Z_{sG} &=& \prod_a \int_{-\infty}^{\infty} d \varphi_a \exp \Bigg( - \sum_a \frac{\left(F_{a\tau}  - \Delta \tau (\mathcal{K}_a - \mathcal{K}_a^\dagger) \right)^2}{2 K_2 \Delta \tau} - \sum_{a, \alpha}\Delta \tau \left[ \frac{K_1 }{2}  F_{a \alpha}^2 + F_{a \alpha} \mathcal{G}_{a \alpha}\right] \nonumber \\
&&~~~~~~~~~~~~~~~~~~~~~~~~~~~+ y \Delta \tau \sum_a \cos(2 \pi (\varphi_a - 2 S \mathcal{Y}_a ))  \Bigg),
\label{sgf}
\eeq
where now
\beq 
F_{a \mu} \equiv \Delta_{\mu} \varphi_a + 2 S  \epsilon_{\mu \nu
\lambda} \Delta_{\nu} {\cal Z}_{i \lambda}.
\eeq
The fixed offsets, ${\cal Y}_a, {\cal{Z}}_i$, are shown pictorially in Fig.~\ref{fig2}. Their values appear to break square lattice
symmetries, but this is only due to a gauge choice: the action above has the full symmetry of the square lattice \cite{zheng89}.

It is now useful to note the following features:
\begin{itemize}
\item
Without the fermion terms, $\mathcal{K}_a$, $\mathcal{G}_{a \alpha}$, 
the partition function $Z_{sG}$ is seen to be
the sine-Gordon field theory of collinear quantum antiferromagnets in 2+1 dimensions \cite{rsprl1,rsprb1}.
\item 
At non-zero temperature, and also without the fermion terms, this reduces to the sine-Gordon model of the
classical dimer model in 2 dimensions \cite{alet05}. There is therefore a phase transition from a confining phase at low $T$
(where $y \rightarrow \infty$), to a deconfined phase at high $T$ (where $y \rightarrow 0$).
\end{itemize}
It is useful to now obtain a theory of a continuous `height' field, $\varphi_a$, coupled to the fermions: this theory has the continuous global symmetry $\varphi_a \rightarrow \varphi_a + c$ for any real constant $c$ when $y=0$. With $y \neq 0$, the symmetry is reduced to a discrete global symmetry, because $c$ has to be an integer. This `shift' symmetry is broken in the confining phase. When shift symmetry remains unbroken in the deconfined
phase where there is a flow to the $y=0$ theory which is dual to non-compact QED. At $T=0$, we expect the shift symmetry to be broken
even in the presence of fermions because integrating out the fermions from the partition function does not introduce any
terms with a new structure in the effective theory of the $\varphi_a$.

To develop more intuition for the above model, one can take a simple mean-field approach and ignore the time-dependent fluctuations of $\varphi_a$. Then we simply have to find the optimum spatial dependence of $\varphi_a$ which will minimize the energy of the following Hamiltonian
\beq
H_{\rm mf} &=& H_f + \sum_{a, \alpha} \left[ \frac{K_1 }{2}  \left(\Delta_{\alpha} \varphi_a + 2 S  \epsilon_{\alpha \nu
\lambda} \Delta_{\nu} {\cal Z}_{i \lambda} \right)^2 + \left(\Delta_{\alpha} \varphi_a + 2 S  \epsilon_{\alpha \nu
\lambda} \Delta_{\nu} {\cal Z}_{i \lambda} \right) \mathcal{G}_{a \alpha}\right] \nonumber\\
&&~~~~~~~~~~~~~~~~ - y \sum_a \cos(2 \pi (\varphi_a - 2 S \mathcal{Y}_a ))
\eeq
The expectation value of $H$ has to be minimized by picking a spatial form for $\varphi_a$ and finding the ground state energy of the fermions in such a background. Without the fermions, this was exactly the procedure followed in Ref.~\cite{rsprb1}, and then it yielded the columnar VBS state.

The explicit form of $H_{{\rm mf}}$ is given by,
\beq
H_{\rm mf} &=& H_f + \sum_{a_x, a_y} \Bigg[ \frac{K_1}{2} \left( \varphi_{a+2 \hat{e}_x} - \varphi_a \right)^2
+ \frac{K_1}{2} \left( \varphi_{a+2 \hat{e}_y} - \varphi_a \right)^2 - y \cos(2 \pi(\varphi_a -2 S \mathcal{Y}_a))
 \nonumber \\
&~&~~+ \lambda (-1)^{a_x + a_y} \left( \varphi_{a + 2 \hat{e}_x} - \varphi_a \right) f_{a + \hat{e}_x - \hat{e}_y,y}^\dagger f_{a + \hat{e}_x - \hat{e}_y,y}^{\vphantom \dagger} \nonumber \\
&~&~~-\lambda (-1)^{a_x + a_y} \left( \varphi_{a + 2 \hat{e}_y} - \varphi_a \right) f_{a + \hat{e}_y - \hat{e}_x,x}^\dagger f_{a + \hat{e}_y - \hat{e}_x,x}^{\vphantom \dagger} \Bigg].
\eeq
(Recall $|\hat{e}_\alpha| = 1/2$.)
We have written out the explicit form of $\mathcal{G}_a$ in terms of the $f$ fermions; the fermions are now expressed in terms of dual lattice
co-ordinates, rather than the direct lattice co-ordinates used earlier (and we have also dropped the spin index on the fermions).
It is easy to see that the $\mathcal{Z}_{i \lambda}$ drop out, as they only couple to the fermions as a chemical potential. We have computed the bond patterns that minimize the energy after integrating out the fermions in the above Hamiltonian; this is safe to do given that the fermions couple to the gradient of the height-field. The details are presented in Appendix \ref{app:u1p}. 

\section{Effective theory for $\mathbb{Z}_2$ FL*}
\label{sec:z2}

The important lesson from Section~\ref{sec:u1} was that the monopole Berry phase term (the $\mathcal{Y}_a$ in $\mathcal{Z}_{sG}$)
remained unchanged from the insulating case. This can ultimately be traced to the parameterization in Eq.~(\ref{fqed}) which allowed us to treat
the fermionic and bosonic dimers in terms of a common U(1) gauge field and a gauge neutral fermion; the latter then did not play a role in the local
constraint in Eq.~(\ref{f1}) which is ultimately responsible for the Berry phases. The gauge neutral fermion only coupled to gauge-invariant combinations
of the gauge field, and the most important was the `dipolar' couping between the electric field $F_{a\alpha}$ and the fermion density $\mathcal{G}_a$.

Turning to the $\mathbb{Z}_2$-FL* case, we begin, as before, by recalling the confinement transition in the insulating case. 
The confinement in the insulator was driven by the condensation of visons (particles carrying $\mathbb{Z}_2$ magnetic flux); the visons
transform projectively under the square lattice space group, and this leads to the appearance of valence bond solid order in 
the confining phase \cite{RJSS91,SSMV99,TSMPA00}.

For the doped case, we can expect from the analysis in Section~\ref{sec:u1} that the projective symmetry group (PSG) of the visons
will remain unchanged from the insulator. Also, we again expect the gauge-neutral fermions couple to the $\mathbb{Z}_2$ electric field only via a dipolar coupling. In such a scenario, we expect that it  is safe to integrate out the fermions completely \cite{Morinari02,TGTS09,Kaul08}; they only serve to renormalize the coefficients of the effective theory for the degrees of freedom associated with the gauge-theory. In this section, we shall consider such an effective theory for a $\mathbb{Z}_2$ FL*, where we already imagine having integrated out the fermions. Our goal is then to study the fate of the ground state upon condensing vortices carrying the $\mathbb{Z}_2$ flux (visons).
\subsection{Lattice Hamiltonian}
\label{sec:lat}
\begin{figure}[!ht]
\includegraphics[height=3.0in]{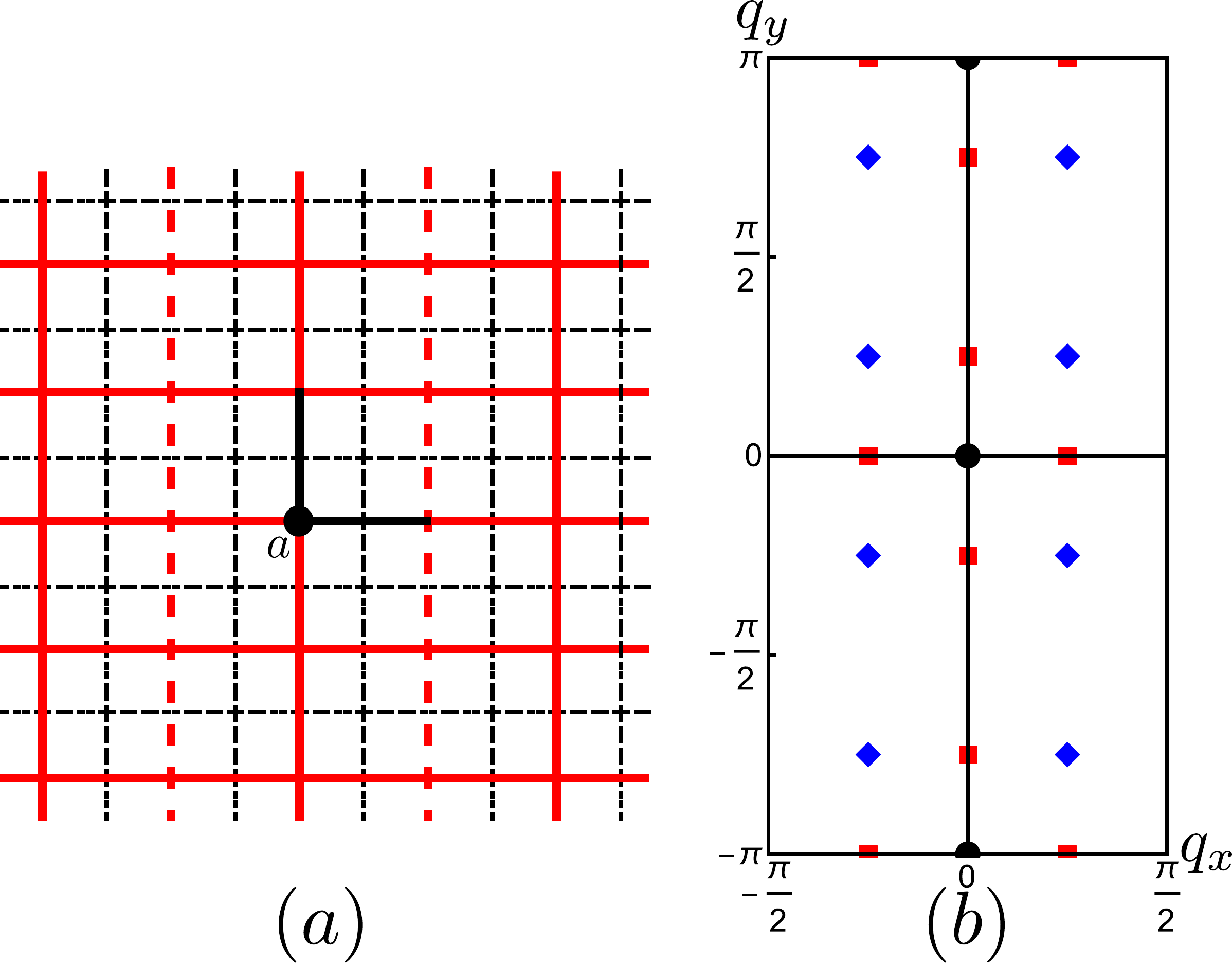}
\caption{(a)~The direct (black) and dual (red) lattices. The dashed dual lattice bonds are frustrated with $J_{ab}=-J$. The dual lattice bonds associated with site $a$ of the dual lattice are shown in black. (b)~Location of the vison dispersion minima in the Brillouin zone in the cases discussed in this work. If $H_1=0$, the minima are located at $(0,0)$ and $(0,\pi)$ (black semi-circles). The blue diamonds represent the diagonal case, and the red squares represent the axial case for $H_1\neq0$.}
\label{fig:latticeminima}
\end{figure}

We represent the $\mathbb{Z}_2$ spin liquid by a fully frustrated Ising model (FFIM) (for a recent derivation, see Ref.~\cite{HPS11} which can be easily adapted to the square lattice) on the dual square lattice \cite{RJSS91,SSMV99,TSMPA00}. The Ising spins ($\sigma^z_a$) represent the vison fields and reside on the sites of the dual lattice with co-ordinates $a = (a_x, a_y)$, with $a_{x,y} \in$ integers (Fig. \ref{fig:latticeminima}(a)). As introduced previously, the vectors $\hat{e}_x = (1/2,0)$ and $\hat{e}_y = (0, 1/2)$.
The bare Hamiltonian is then given by,
\beq
H_0 = - \sum_{\langle a,b \rangle} J_{ab} \, \sigma^z_{a} \sigma^z_{b}
\label{eq:h0} 
\eeq
where $\langle a,b \rangle$ represents nearest-neighbor pairs, and the Ising interaction is uniformly frustrated
with $|J_{ab}| = J$, satisfying the constraint
\beq
\prod_{\square} J_{ab} = - J^4.
\eeq
This Ising model was studied by Villain \cite{Villain77}. For the remainder of this work, we will choose a gauge in which alternating rows of vertical bonds are frustrated, indicated by the dashed lines in Fig.~\ref{fig:latticeminima} (our results are independent of gauge choice), i.e.
\beq
J_{ab} = J(\delta_{b_x,a_x\pm1}\delta_{b_y,a_y}+(-1)^{a_x}\delta_{b_x,a_x}\delta_{b_y,a_y\pm1}).
\eeq
Our goal is to study this model with simple additional couplings allowed under the projective symmetry group (PSG) and investigating the resulting density-wave ground states with non-trivial form-factors.  

We will now add additional two-spin couplings to Eq.~(\ref{eq:h0}): In general, we should consider all terms consistent with the PSG, as in Ref.~\cite{HPS11}.
The PSG transformations corresponding to the different dual lattice symmetries $\mathcal{O}$ of the lattice Ising variables in our gauge choice are summarized below. The transformations centered on the dual lattice sites include: Translation along $x, y$ ($T_{x,y}$), inversion about $x, y$ axes ($I_{x,y}$), and rotation by $\pi/2$ ($R_{\pi/2}$). 

\begin{align}
&T_x : \sigma^z_a \rightarrow (-1)^{a_y}\sigma^z_{a-2\hat{e}_x},~T_y : \sigma^z_a \rightarrow \sigma^z_{a-2\hat{e}_y}, \nonumber \\
&I_x : \sigma^z_{a_x,a_y} \rightarrow \sigma^z_{a_x,-a_y},~I_y : \sigma^z_{a_x,a_y} \rightarrow \sigma^z_{-a_x,a_y}, \nonumber \\
&R_{\pi/2} : \sigma^z_{a_x,a_y} \rightarrow  (-1)^{a_x a_y} \sigma^z_{a_y,-a_x}.
\label{eq:psgr}
\end{align} 
The Hamiltonian must be invariant under these transformations, after also applying the symmetry operations to the coupling constants. The simplest allowed two-spin couplings that have the same sign on all bonds are (note $|\hat{e}_i | = 1/2$)
\beq
H_ 1 &=& J_1 \sum_{a} \left[ \sigma^z_{a} \sigma^z_{a + 4 \hat{e}_x} + \sigma^z_{a} \sigma^z_{a + 4 \hat{e}_y} \right] \nonumber\\
&+& J_2 \sum_{a} \left[ \sigma^z_{a} \sigma^z_{a + 4 \hat{e}_x +  4 \hat{e}_y} + \sigma^z_{a} \sigma^z_{a +  4 \hat{e}_x -  4 \hat{e}_y } \right]\nonumber\\
 &+&  J_3 \sum_{a} \left[ \sigma^z_{a} \sigma^z_{a + 8 \hat{e}_x} + \sigma^z_{a} \sigma^z_{a + 8 \hat{e}_y} \right]. 
\label{Hp}
\eeq

We shall now be interested in studying ground states of $H=H_0+H_1$ in Eqs.~(\ref{eq:h0}) and (\ref{Hp}) in the following sections. Notice that, as written, $H$ has no dynamics. However, we'll explicitly include the kinetic-energy, that descends from a transverse-field term, in subsequent sections. 

\subsection{Continuum Field Theory}

We take the continuum limit of the model defined in $H_0+H_1$ in Eqs.~(\ref{eq:h0}) and (\ref{Hp}) as in Ref.~\cite{BBBSS05}: This can be done by softening the Ising spins ($\sigma^z_a\rightarrow\phi_a\in\mathbb{R}$). In momentum space, $q=(q_x,q_y)$, we define
\beq
\phi_q = \frac{1}{(L_x L_y)^{1/2}}\sum_a \phi_a e^{-iq\cdot a}
\eeq  
and introduce $\Phi_q^\dagger= (\phi_q^*~~\phi^*_{q+K_x})$, where $K_x=(\pi,0)$ \cite{HPS11}. We thus get 
\beq
H &=& \sum_q \Phi_q^\dagger H(q) \Phi_q,~\tn{where}\\
H(q) &=& -\xi_0(q_x) \tau_z - \xi_0(q_y) \tau_x + \xi_1(q) \tau_0,\\
 \xi_0(q_i) &=& 2J\cos(q_i)~~(i=x, y),\\
 \xi_1(q) &=& 2J_1\left(\cos 2q_x+\cos 2q_y\right) \nonumber\\
 &+& 2J_2\left(\cos (2q_x+2q_y)+\cos (2q_x-2q_y)\right) \nonumber\\
 &+& 2J_3\left(\cos 4q_x+\cos 4q_y\right),
\eeq
and the Brillouin zone is defined as $-\pi/2 < q_x\leq\pi/2,-\pi < q_y\leq \pi$. The vison dispersion is thus
\begin{align}
&\xi^\pm(q) = \xi^\pm_0(q) + \xi_1(q), \nonumber \\
&\xi^\pm_0(q)=\pm \sqrt{\xi_0^2(q_x)+\xi_0^2(q_y)}.
\label{eq:vdisp}
\end{align}
See Appendix \ref{em} for the expressions of the corresponding eigenmodes, $v^\pm(q)$, and their transformation rules under the PSG.

If we set $H_1=0$, the lower band, $\xi^-(q)$, of the vison dispersion has minima at $(0,0)$ and $(0,\pi)$ in our gauge choice. In order to obtain bond-density waves (BDW) with incommensurate wavevectors, the dispersion minima need to be located at points other than $(0,0)$ and $(0,\pi)$. We must thus adjust $H_1$ to obtain such a scenario. The simplest possible cases consistent with all of the symmetries mentioned above are (see Fig. \ref{fig:latticeminima}(b)):
\begin{itemize} 
\item The diagonal case with degenerate global minima at $q^{(n)} = (\pm p, \pm p), (\pm p,\pi \pm p)$. This is realized when, for example, $1/(4\sqrt{2}) < J_1/J$ and $J_2 <2J_3$ (all $J_i$'s$>0$). 
\item The axial case with degenerate global minima at $q^{(n)} = (\pm p, 0), (0, \pm p), (\pm p, \pi), (0, \pi\pm p)$. This can be realized when, for example, $1/(4\sqrt{2}) < J_1/J < 1/4$ and $2J_3-J_1/2+J/8>J_2 > 2J_3$ (all $J_i$'s$>0$). For $J_2$ very large, $p$ will be pinned to $\pi/2$.
\end{itemize} 

\begin{figure*}[!ht]
\includegraphics[height=3.0in]{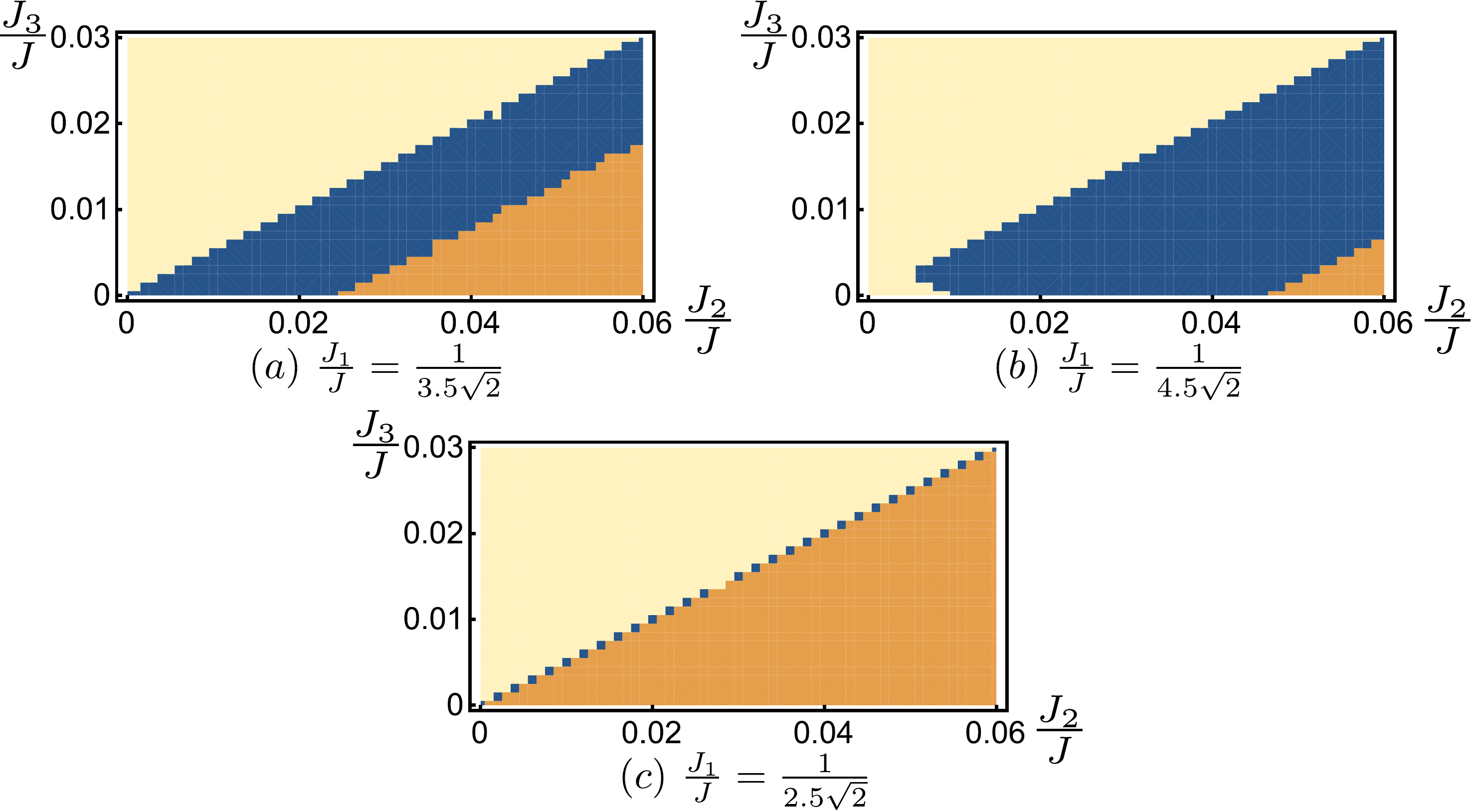}
\caption{Phase diagram showing the different types of dispersion minima for various values of $J_1,J_2,J_3$. The blue regions correspond to the axial case discussed above, and the off-white regions correspond to the diagonal case. In the orange regions, at least one of the two momentum coordinates of the minima is $\pi/2$. If $J_1/J>1/4$, the axial case does not exist for any values of $J_2,J_3$. On the boundaries of the blue and off-white regions, the minimum of the dispersion lies on a contour passing through both axial and diagonal points.}
\label{fig:pd1}
\end{figure*}

The full $J_1, J_2, J_3$ phase diagram is complicated, and cross sections are illustrated in Fig. \ref{fig:pd1}. The value of $p$ and the depth of the minima is controlled by the ratios $J_{i}/J$.

In addition, it is possible to have other cases, but they lead to more minima and we hence refrain from discussing them here {\footnote{One such example is when $J_1/J > 1/(4\sqrt{2})$ and $J_2 = J_3 = 0$, which makes the minimum of the dispersion lie on a contour, leading to an infinite number of degenerate minima.}}. Note that it is also possible to include two-spin couplings that do not have the same sign on all bonds in $H_1$, but this doesn't change the fact that the simplest possible configurations of incommensurate minima that can be obtained are the axial and diagonal cases discussed above, which is just a consequence of the symmetries of the problem.

The real space magnetization $\phi_a$ may be expressed by associating complex amplitudes $\varphi\left(q^{(n)}\right)=\varphi^\ast\left(-q^{(n)}\right)$ with the different minima of the lower band located at $q^{(n)}$. The eigenmodes corresponding to the minima then realize representations $\mathcal{O}_{nm}$ of the PSG, which, for incommensurate minima, are isomorphic to representations of the symmetry group of the square lattice with $\pi$ flux per plaquette and magnetic, instead of regular, translations, i.e.
\beq
\phi_a &=& \sum_n \varphi\left(q^{(n)}\right)v^-\left(q^{(n)}\right)\\
\mathcal{O}(\phi_a) &=& \sum_{nm} \varphi\left(q^{(n)}\right)\mathcal{O}_{nm}v^-\left(q^{(m)}\right)
\label{eq:mag}
\eeq
We label the amplitudes $\varphi\left(q^{(n)}\right)$ in the axial and diagonal cases using complex fields as in Table~(\ref{tab:fieldlabels}). 
\begin{table*}[ht]
\centering
\begin{tabular}{||c | c | c||} 
\hline
Field & Axial & Diagonal \\ [0.5ex] 
\hline\hline
$\varphi_1$ & $\varphi(p,0)$ & $\varphi(p,p)$  \\ 
$\varphi_2$ & $\varphi(0,-p)$ & $\varphi(p,-p)$ \\
$\varphi_3 $& $\varphi(p,\pi)$ & $\varphi(p,-\pi+p)$ \\
$\varphi_4$ & $\varphi(0,\pi-p)$ & $\varphi(p,\pi-p)$ \\
\hline
\end{tabular}
\caption{Complex fields corresponding to the different dispersion minima of the soft-spin Ising fields. The complex conjugate of a given field naturally corresponds to the minimum with the opposite momentum.}
\label{tab:fieldlabels}
\end{table*}
The PSG operations are then given by Table~(\ref{tab:PSG}).
\begin{center}
\begin{table*}[ht]
\begin{tabular}{|l|l|l|l|l|l|l|l|l|}
\hline
              & \multicolumn{2}{l|}{$\varphi_1$}                                                & \multicolumn{2}{l|}{$\varphi_2$}                                                                    & \multicolumn{2}{l|}{$\varphi_3$}                                                & \multicolumn{2}{l|}{$\varphi_4$}                                                                    \\ \hline
$\mathcal{O}$ & Axial                                  & Diagonal                               & Axial                                            & Diagonal                                         & \multicolumn{1}{c|}{Axial}             & Diagonal                               & Axial                                            & Diagonal                                         \\ \hline
$T_x$         & $e^{-ip}\varphi_3$                     & $e^{-ip}\varphi_3$                     & $\varphi_4$                                      & $e^{-ip}\varphi_4$                               & $e^{-ip}\varphi_1$                     & $e^{-ip}\varphi_1$                     & $\varphi_2$                                      & $e^{-ip}\varphi_2$                               \\
$T_y$         & $\varphi_1$                            & $e^{-ip}\varphi_1$                     & $e^{ip}\varphi_2$                                & $e^{ip}\varphi_2$                                & -$\varphi_3$                           & -$e^{-ip}\varphi_3$                    & $-e^{ip}\varphi_4$                               & $-e^{ip}\varphi_4$                               \\
$R_{\pi/2}$   & $\frac{\varphi_2+\varphi_4}{\sqrt{2}}$ & $\frac{\varphi_2+\varphi_4}{\sqrt{2}}$ & $\frac{\varphi_1^\ast+\varphi_3^\ast}{\sqrt{2}}$ & $\frac{\varphi_1^\ast+\varphi_3^\ast}{\sqrt{2}}$ & $\frac{\varphi_2-\varphi_4}{\sqrt{2}}$ & $\frac{\varphi_2-\varphi_4}{\sqrt{2}}$ & $\frac{\varphi_1^\ast-\varphi_3^\ast}{\sqrt{2}}$ & $\frac{\varphi_1^\ast-\varphi_3^\ast}{\sqrt{2}}$ \\
$I_x$         & $\varphi_1$                       & $\varphi_2$                       & $\varphi_2^\ast$                                      & $\varphi_1$                                 & $\varphi_3$                       & $\varphi_4$                       & $\varphi_4^\ast$                                      & $\varphi_3$                                 \\
$I_y$         & $\varphi_1^\ast$                            & $\varphi_2^\ast$                            & $\varphi_2$                                 & $\varphi_1^\ast$                                      & $\varphi_3^\ast$                            & $\varphi_4^\ast$                            & $\varphi_4$                                 & $\varphi_3^\ast$                                      \\ \hline
\end{tabular}
\caption{PSG transformations of the complex fields at the minima.}
\label{tab:PSG}
\end{table*}
\end{center}

\subsection{Low energy field theory}
We can now write down the action for the low energy theory of these complex fields by considering the most general real polynomials in the fields that are invariant under the transformations in table (\ref{tab:PSG}), and under $\varphi_n\rightarrow-\varphi_n$. We restrict ourselves to up to quartic terms in the fields as this shall prove sufficient to break all continuous symmetries apart from certain unbreakable $U(1)$ symmetries associated with rotating the phases of the complex fields. These unbreakable symmetries are a consequence of the required insensitivity of the Lagrangian to the incommensurate phases acquired by the $\varphi_n$ under translations. 

We obtain, for the diagonal case, the Lagrangian density
\beq
\mathcal{L}_d &=& \sum_{n=1}^4\left(|\partial_\tau\varphi_n|^2 + \frac{K}{2}|\nabla_n\varphi_n|^2 + \frac{r}{2}|\varphi_n|^2\right) + \frac{U}{4}\left(\sum_{n=1}^4|\varphi_n|^2\right)^2+\frac{W}{6}\left(\sum_{n=1}^4|\varphi_n|^2\right)^3 \nonumber \\
&+& \frac{V_0}{4}\left((\varphi_1 \varphi_3^\ast-\varphi_1^\ast \varphi_3)^2+(\varphi_2 \varphi_4^\ast-\varphi_2^\ast \varphi_4)^2\right) + V_1(|\varphi_1|^2 + |\varphi_3|^2)(|\varphi_2|^2 + |\varphi_4|^2)\nonumber \\
&+& V_2\bigg[(\varphi_1 \varphi_3^\ast + \varphi_1^\ast \varphi_3) (\varphi_2 \varphi_4^\ast + \varphi_2^\ast \varphi_4)  - 2 (|\varphi_1|^2 |\varphi_4|^2 + |\varphi_2|^2 |\varphi_3|^2)\bigg],
\label{eq:ld}
\eeq
where we also added the $O(8)$ symmetric 6th order term to ensure convexity of the free energy for any set of values of the quartic couplings. The gradient terms may be anisotropic but must transform appropriately under rotation and inversions. 

In the axial case, an additional set of terms is allowed, which would break inversion symmetry, if included in the diagonal case.
\beq
\mathcal{L}_a=\mathcal{L}_d &+& V_3\bigg[(|\varphi_1|^2-|\varphi_3|^2)^2 - (|\varphi_2|^2 - |\varphi_4|^2)^2 + (\varphi_2 \varphi_4^\ast+\varphi_2^\ast \varphi_4)^2-(\varphi_1 \varphi_3^\ast+\varphi_1^\ast \varphi_3)^2\bigg].
\label{eq:la}
\eeq
The symmetry of $\mathcal{L}_a$ is reduced from $O(8)$ to $\mathbb{Z}_4\times\mathbb{Z}_2\times\mathbb{Z}_2\times\mathbb{Z}_2\times U(1)\times U(1)$ by the $V$ couplings, and that of $\mathcal{L}_d$ is $\mathbb{Z}_4\times\mathbb{Z}_2\times\mathbb{Z}_2\times\mathbb{Z}_2\times\mathbb{Z}_2\times  U(1)\times U(1)$. 

\subsection{Density wave observables}
\label{sec:dwo}

Finally we have to address the issue of the observables for density-wave order. These are defined on the \textit{direct} lattice bonds, so the observables for the direct lattice bonds pointing in the $x$ direction (which we will call $\rho_x$) correspond to the dual lattice bonds intersected by them, which point in the $y$ direction, and vice-versa.  The dual lattice bonds associated with a dual lattice site are defined to be the ones pointing outwards from it in the positive $x$ and $y$ directions (See Fig. \ref{fig:latticeminima}(a)), with direct lattice bond density observables $\rho_y^+$ and $\rho_x^+$ respectively ($\rho_y^-$ and $\rho_x^-$ are the observables on the bonds pointing outwards in the negative $x$ and $y$ directions) . The bond density observables must be real and quadratic in the complex fields. We can express them as 
\begin{align}
&\rho_{x,y}^+(a) = \sum_n e^{i\tilde{q}^{(n)}\cdot a}\rho_{x,y}^+(\tilde{q}^{(n)}), \nonumber \\
&\left\{\tilde{q}^{(n)}\right\}=\left\{q^{\prime(j)}+q^{\prime(k)}\right\},~~~q^{\prime(j)},q^{\prime(k)}\in \left\{q^{(n)}\right\}\cup\left\{q^{(n)}+K_x\right\}, \nonumber \\
&\rho_{x,y}^-(a)=\rho_{x,y}^+(a-2\hat{e}_{y,x}) = \sum_n e^{i\tilde{q}^{(n)}\cdot a}\rho_{x,y}^-(\tilde{q}^{(n)}).
\label{eq:rho}
\end{align}

To better understand what the density observables actually represent, we can imagine coupling our extended FFIM to fermionic dimers $f$ living on the direct lattice bonds, as introduced earlier. Clearly, the simplest Ising operator with the right symmetries for the fermions to couple to is the bond energy itself, and so we have, in the conventions of Fig. \ref{fig:latticeminima}(a) (fermion spin indices are dropped as the Hamiltonian is diagonal in them),
\begin{align} 
&H_{If} = - \lambda \sum_a \left(E^+_x(a) f^\dagger_{a+\hat{e}_x-\hat{e}_y,y} f_{a+\hat{e}_x-\hat{e}_y,y}^{\vphantom \dagger}+E^+_y(a) f^\dagger_{a+\hat{e}_y-\hat{e}_x,x} f_{a+\hat{e}_y-\hat{e}_x,x}^{\vphantom \dagger}\right), \nonumber \\
&E^\pm_x(a) = J_{a,a\pm2\hat{e}_x}\phi_a\phi_{a\pm2\hat{e}_x},~E^\pm_y(a) = J_{a,a\pm2\hat{e}_y}\phi_a\phi_{a\pm2\hat{e}_y},
\end{align}
and the bond observables are simply the dimer densities on the bonds $\rho^+_x(a)=f^\dagger_{a+\hat{e}_y-\hat{e}_x,x}f_{a+\hat{e}_y-\hat{e}_x,x}$, $\rho^+_y(a)=f^\dagger_{a+\hat{e}_x-\hat{e}_y,y}f_{a+\hat{e}_x-\hat{e}_y,y}$. Defining $\Psi^\dagger_a=(f^\dagger_{a+\hat{e}_y-\hat{e}_x,x},~f^\dagger_{a+\hat{e}_x-\hat{e}_y,y})$ as in Ref. \cite{PAS15}, the momentum space Lagrangian density for the dimers in the dilute limit has the generic form
\beq
\mathcal{L}_f = \Psi^\dagger(k)G_0^{-1}(k)\Psi(k) + \mathcal{L}_{\rm int}(\Psi,\Psi^\dagger) \rightarrow  \Psi^\dagger(k)G^{-1}(k)\Psi(k),
\eeq
where $\mathcal{L}_{\rm int}(\Psi,\Psi^\dagger)$ is an unspecified interaction term and $k = (\mathbf{k},i\omega_n)$.

Integrating out the fermions allows us to generate an expression for the dimer density on a given bond. We obtain the generic expression
\begin{align}
&\rho_{j=x,y}^\pm(a) = \rho_0 + \lambda \sum_{a^\prime,l={x,y}} E_l^\pm(a^\prime)\sum_{k,q}\mathrm{Tr}\left[G(k)M_l G(q)M_j\right]e^{i(a^\prime-a)\cdot(\mathbf{k}-\mathbf{q})}, \nonumber \\
&M_x = \frac{1}{2}(\tau_0-\tau_z),~~M_y = \frac{1}{2}(\tau_0+\tau_z).
\end{align}
This turns out to be a positively weighted linear combination of energies of dual lattice bonds intersected by nearby bonds, with the highest weight going to the dual lattice bond intersecting the pertinent bond itself. Thus, in the dilute limit, it is a good approximation to take the bond observables to be the bond energies of the intersecting dual lattice bonds. 

In general, the $\rho_{x,y}$ are quadratic in the $\varphi_n$ and obey the following transformation rules \cite{BBBSS05}
\beq
&&T_x:~\rho_{x,y}^\pm(\tilde{q}^{(n)}_x,\tilde{q}^{(n)}_y)\rightarrow e^{-i\tilde{q}^{(n)}_x}\rho_{x,y}^\pm(\tilde{q}^{(n)}_x,\tilde{q}^{(n)}_y),~ T_y:~\rho_{x,y}^\pm(\tilde{q}^{(n)}_x,\tilde{q}^{(n)}_y)\rightarrow e^{-i\tilde{q}^{(n)}_y}\rho_{x,y}^\pm(\tilde{q}^{(n)}_x,\tilde{q}^{(n)}_y), \nonumber \\
&&I_x:~\rho_{x}^\pm(\tilde{q}^{(n)}_x,\tilde{q}^{(n)}_y)\rightarrow \rho_{x}^\mp(\tilde{q}^{(n)}_x,-\tilde{q}^{(n)}_y),~\rho_{y}^\pm(\tilde{q}^{(n)}_x,\tilde{q}^{(n)}_y)\rightarrow \rho_{y}^\pm(\tilde{q}^{(n)}_x,-\tilde{q}^{(n)}_y), \nonumber \\
&&I_y:~\rho_{x}^\pm(\tilde{q}^{(n)}_x,\tilde{q}^{(n)}_y)\rightarrow \rho_{x}^\pm(-\tilde{q}^{(n)}_x,\tilde{q}^{(n)}_y),~\rho_{y}^\pm(\tilde{q}^{(n)}_x,\tilde{q}^{(n)}_y)\rightarrow \rho_{y}^\mp(-\tilde{q}^{(n)}_x,\tilde{q}^{(n)}_y),  \nonumber \\
&&R_{\pi/2}:~\rho_{x}^\pm(\tilde{q}^{(n)}_x,\tilde{q}^{(n)}_y)\rightarrow \rho_{y}^\pm(\tilde{q}^{(n)}_y,-\tilde{q}^{(n)}_x),~\rho_{y}^\pm(\tilde{q}^{(n)}_x,\tilde{q}^{(n)}_y)\rightarrow \rho_{x}^\mp(\tilde{q}^{(n)}_y,-\tilde{q}^{(n)}_x).
\label{eq:denstrans}
\eeq
So,
\ben
\rho_{x,y}^\pm(\tilde{q}^{(n)}_x,\tilde{q}^{(n)}_y) = S_{x,y}^\pm(\tilde{q}^{(n)}_x,\tilde{q}^{(n)}_y)f^{\varphi\varphi}_{x,y}(\tilde{q}^{(n)}_x,\tilde{q}^{(n)}_y),
\label{eq:formfactor}
\een
where the $S$ are form factors that cannot be determined by symmetry considerations and $f^{\varphi\varphi}_{x,y}(\tilde{q}^{(n)}_x,\tilde{q}^{(n)}_y)$ are quadratic polynomials in the $\varphi_n$. The transformation of the $f$'s is effected by just transforming the complex fields they depend on according to Table~\ref{tab:PSG}. In addition, the form factors $S$ are smooth complex functions of their arguments, whose details depend upon the exact definition of the bond density observables, and satisfy $S^{\pm\ast}_{x,y}(-\tilde{q}^{(n)}_x,-\tilde{q}^{(n)}_y)=S^{\pm}_{x,y}(\tilde{q}^{(n)}_x,\tilde{q}^{(n)}_y)$. In order for Eq.~(\ref{eq:denstrans}) to hold, the $S$'s must satisfy the following additional constraints:
\beq
&&S^\pm_{x}(\tilde{q}^{(n)}_x,\tilde{q}^{(n)}_y) = S^\pm_{x}(-\tilde{q}^{(n)}_x,\tilde{q}^{(n)}_y) =  S^\mp_{x}(\tilde{q}^{(n)}_x,-\tilde{q}^{(n)}_y), \nonumber \\
&&S^\pm_{y}(\tilde{q}^{(n)}_x,\tilde{q}^{(n)}_y) = S^\pm_{y}(\tilde{q}^{(n)}_x,-\tilde{q}^{(n)}_y) =  S^\mp_{y}(-\tilde{q}^{(n)}_x,\tilde{q}^{(n)}_y), \nonumber \\
&&S^\pm_{x}(\tilde{q}^{(n)}_x,\tilde{q}^{(n)}_y) = S^\pm_{y}(\tilde{q}^{(n)}_y,-\tilde{q}^{(n)}_x) = S^\mp_{x}(-\tilde{q}^{(n)}_x,-\tilde{q}^{(n)}_y).
\label{eq:ffconst}
\eeq

\section{Results}
\label{sec:results}

\subsection{Ground states}
\label{sec:gs}
We now minimize the free energy associated with Eqs.~(\ref{eq:ld}) and (\ref{eq:la}). Expressing $\varphi_n=\psi_n e^{i\theta_n}$, we obtain the free energy density $\mathcal{F}_a =\mathcal{L}_a - \sum_{n=1}^4 |\partial_\tau\varphi_n|^2$
\beq
\mathcal{F}_a &=& \frac{K}{2}|\nabla_n\varphi_n|^2 + \frac{r}{2} \left(\sum_{i=1}^4 \psi_i^2\right)+\frac{U}{4} \left(\sum_{i=1}^4 \psi_i^2\right)^2 + \frac{W}{6} \left(\sum_{i=1}^4 \psi_i^2\right)^3  \nonumber \\
&-& V_0 \bigg[\psi_1^2 \psi_3^2 \sin ^2(\theta_{13})+\psi_2^2 \psi_4^2 \sin ^2(\theta_{24})\bigg]\nonumber\\
&+& V_1\bigg[\left(\psi_1^2+\psi_3^2\right) \left(\psi_2^2+\psi_4^2\right)\bigg] \nonumber \\
&-&2V_2 \bigg[\psi_1^2 \psi_4^2+\psi_2^2 \psi_3^2-2 \psi_1 \psi_2 \psi_3 \psi_4 \cos (\theta_{13}) \cos (\theta_{24})\bigg]\nonumber \\
&+&V_3\bigg[\psi_1^4 + \psi_3^4 - \psi_2^4 - \psi_4^4 - 4 \psi_1^2 \psi_3^2 + 4 \psi_2^2 \psi_4^2\nonumber\\
&&-2 \psi_1^2 \psi_3^2 \cos(2\theta_{13}) + 2 \psi_2^2 \psi_4^2 \cos(2\theta_{24})\bigg],
\label{eq:lgfe}
\eeq
where $\theta_{13}=\theta_1-\theta_3,~\theta_{24}=\theta_2-\theta_4 $.

\begin{figure*}[!ht]
\includegraphics[height=2.5in]{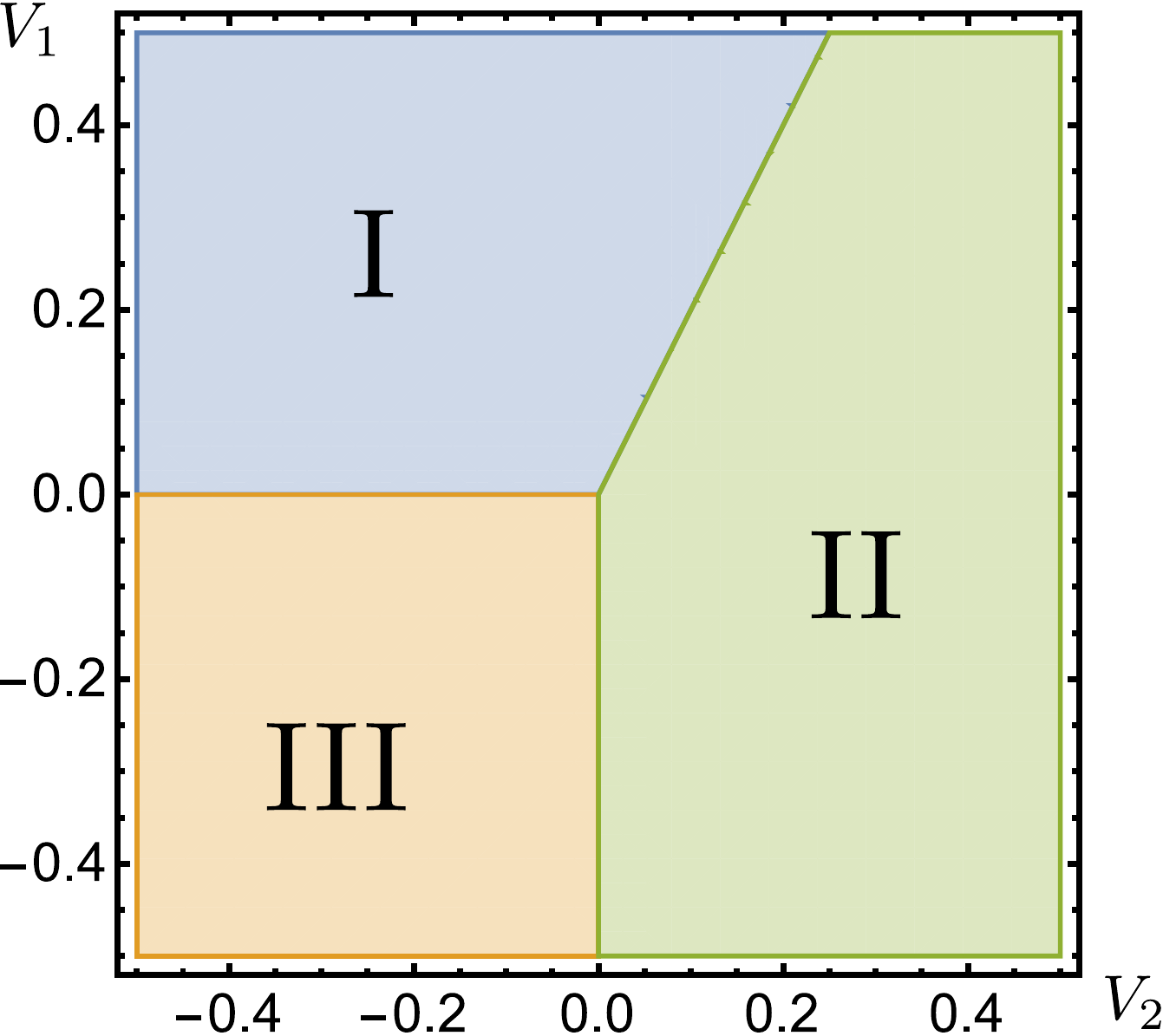}
\caption{The phase diagram of Eq.~(\ref{eq:lgfe}) in the region of parameter space given by $r<0,~U,W>0,~V_0<0,~V_3=0$, and hence applicable to both the axial and diagonal cases. The phases are described in the text.}
\label{fig:lgpd}
\end{figure*}

We are interested in condensing visons, i.e. $r<0$, and we require $K>0$, $U>0$, $W>0$ for thermodynamic stability. In order to get nontrivial minima, we minimize $\mathcal{F}_a$ for different choices of $V_0,~V_1,~V_2,~V_3$. Moreover, for the diagonal case, we can set $V_3=0$. The phase diagram is shown in Fig. \ref{fig:lgpd}, generated for various values of $V_1,~V_2$ for $V_0<0$, $V_3=0$ and remaining parameters as above. 

The different regions plotted in the phase-diagram have the following properties:
\begin{itemize}
\item In the phase labeled by I, the ground states are given by
\beq
\psi_1^2+\psi_3^2 &=& \frac{\sqrt{U^2-4Wr}-U}{2 W},~~\theta_{13} = 0,\nonumber\\
\psi_2 =\psi_4 &=& 0.
\eeq
The state above is degenerate to the one obtained by $\{1,3\}\leftrightarrow\{2,4\}$. 

In the phase above, if we additionally allow for $V_3\neq0$ (i.e. the axial case), then $V_3>0$ gives a set of degenerate minima specified by
\beq
\psi_1&=&\psi_3(=\alpha)\neq0,~\theta_{13}=0,~\psi_2=\psi_4=0\nonumber\\
\psi_1&=&\psi_3(=\alpha)\neq0,~\theta_{13}=\pi,~\psi_2=\psi_4=0,\nonumber \\ 
\psi_1&=&\psi_3=0,~\psi_4=0,~\psi_2(=\sqrt{2}\alpha)\neq0,\nonumber\\
\psi_1&=&\psi_3=0,~\psi_2=0,~\psi_4(=\sqrt{2}\alpha)\neq0.
\eeq 
$V_3<0$ gives the same configurations with $\{1,3\}\leftrightarrow\{2,4\}$. 

On the other hand, for $V_0>0$, we get the same result as long as $|V_3|>V_0/4$. For $|V_3|<V_0/4$ we instead get $\theta_{13}=\pm \pi/2$ instead of $\pi$. 

\item In phase II, the ground states have either $\psi_1=\psi_4$ or $\psi_2=\psi_3$ (degenerate) for $V_0<0$ and $V_3=0$. The continuous degeneracy of the ground states is $U(1)\times U(1)$. Since the nonzero complex fields are associated with different incommensurate wavevectors, there will be simultaneous extra modulation at more than one wavevector. 

\item In phase III, again for $V_0<0$ and $V_3=0$, all the $\psi$'s are nonzero in the ground states. 
\end{itemize}
Phase I is the most interesting phase; the ground state has modulations of the condensed vison-fields at wavevectors $\pm q^{(n)}$ and $\pm q^{(n)}+(0,\pi)$. This is is the closest scenario to the pattern observed in the experiments on the underdoped cuprates, as we shall discuss below. Moreover, the axial case in this phase has ordering wavevectors in the experimentally observed directions. Thus we shall mainly study the features associated with phase I in this work. There is also a continuous $U(1)$ degeneracy of the ground states coming from the freedom to choose different arguments of the nonzero complex fields.

We also performed a one-loop renormalization group analysis of the theories defined by Eqs.~(\ref{eq:ld}) and (\ref{eq:la}). We find the same set of eight nontrivial critical fixed points in both the axial and diagonal cases, out of which one is the $O(8)$ Wilson-Fisher fixed point with $U\neq0$ and $V_i=0$. This fixed point is stable as long as all the $V_i$ are zero. The seven remaining fixed points also have $V_3=0$, however, they are all unstable and the couplings flow to infinity when they are displaced slightly from their fixed point values. It would be interesting to see if stable fixed points emerge at higher loop orders.

\subsection{Bond patterns}
\begin{figure*}[!ht]
\includegraphics[height=2.25in]{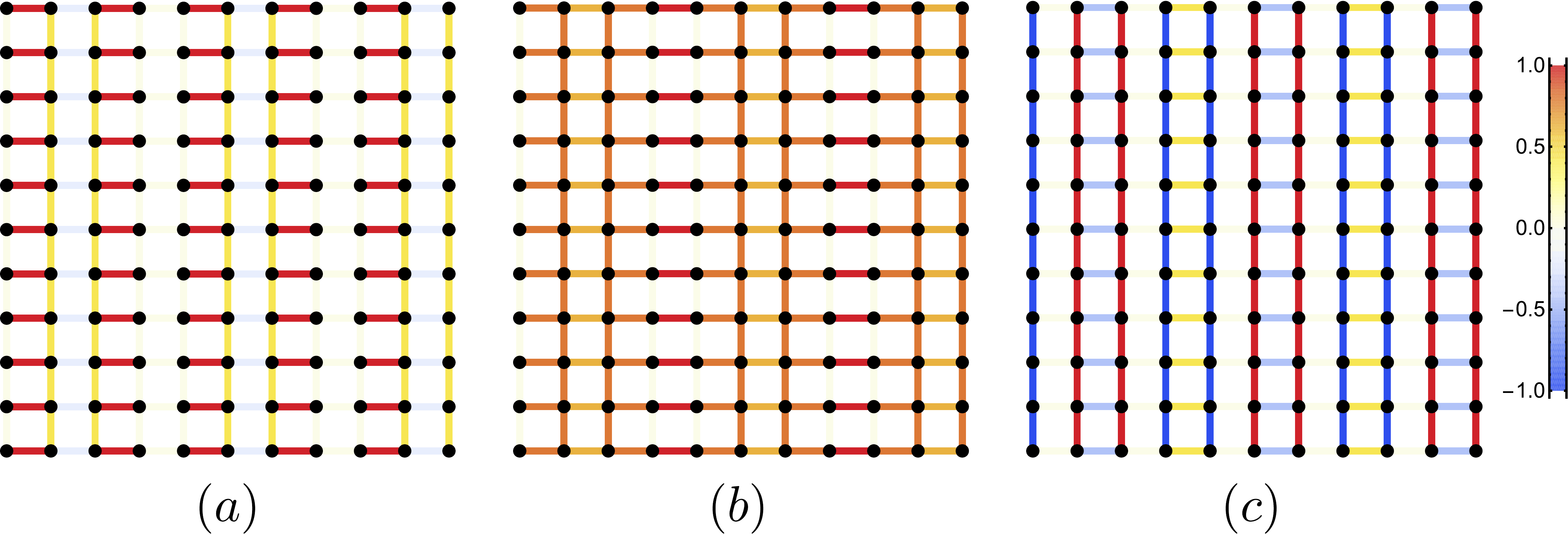}
\caption{Individually normalized bond density wave components on the direct lattice using observables of Eq.~(\ref{eq:rhoe}) in the region I of Fig. \ref{fig:lgpd} with $V_3<0$. We chose the axial case with $p\approx\pi/4$, and the Landau minimum specified by $\psi_1=1.0,~\psi_3=0,~\theta_1=3\pi/4,~\psi_2=\psi_4=0$. (a) The BDW pattern with all its constituent wavevectors, i.e, $(0,0)$,~$(\pm\pi/2,0)$,~$(\pi,0)$. (b) The pattern with component with wavevector $(\pi,0)$ removed. (c) The component with wavevectors $(\pm\pi/2,0)$, showing 72.5\% d and 27.5\% s' form factors. Degenerate ground states produce patterns rotated by $90^\circ$ about the dual lattice sites.}
\label{fig:patterns1}
\end{figure*}

As highlighted in Sec.~\ref{sec:dwo}, there is a certain degree of ambiguity in choosing the appropriate gauge-invariant observable associated with the density wave. A natural choice for the density wave observables would be the bond energies of the dual lattice bond intersecting the specified direct lattice bond, i.e. 
\beq
\textnormal{Choice A :} \begin{cases}\rho^\pm_x(a) = E^\pm_y(a) = J_{a,a\pm2\hat{e}_y}\phi_a\phi_{a\pm2\hat{e}_y},\\
\rho^\pm_y(a) = E^\pm_x(a) = J_{a,a\pm2\hat{e}_x}\phi_a\phi_{a\pm2\hat{e}_x}.\end{cases}
\label{eq:rhoe}
\eeq
An equally acceptable choice, especially in a regime of strong ``dimer" interactions, involves taking the bond observables to be certain linear combinations of energies of nearby bonds allowed by symmetry, i.e.
\beq
\textnormal{Choice B :} \begin{cases} \rho^\pm_x(a) = E^\pm_y(a) + \frac{E^\pm_y(a-2\hat{e}_x)+E^\pm_y(a+2\hat{e}_x)}{2},\\
\rho^\pm_y(a) = E^\pm_x(a) + \frac{E^\pm_x(a-2\hat{e}_y)+E^\pm_x(a+2\hat{e}_y)}{2}.
\end{cases}
\eeq
The rationale for choice B will become clear below. We evaluate the densities using Eq.~(\ref{eq:mag}). The bond patterns on the direct lattice are shown in Fig. \ref{fig:patterns1} for a particular set of values in parameter space in the axial case.

The $\pi$ flux per plaquette for the visons invariably results in modulation wavevectors appearing both near $\Q=(0,0)$ (i.e. a nematic component) and $\Q=(\pi,0),~(0,\pi)$ in general. Working with choice A gives rise to patterns like the one shown in Fig. \ref{fig:patterns1}(a), which has in addition to the above components, a component along $\Q_a=(\pm\pi/2,0)$; this is determined by the value of $p~(\approx\pi/4)$. There is also a continuous sliding symmetry for the bond density waves arising from the continuous $U(1)$ degeneracy of the ground states. 

If instead, we work with choice B, the extra wavevectors near $(\pi,0),~(0,\pi)$ can be removed. Fig. \ref{fig:patterns1}(b) shows such a pattern, with a uniform nematic component and a density-wave with $\Q_a$. Fig. \ref{fig:patterns1}(c) shows the density wave in Fig.~\ref{fig:patterns1}(b), with the nematic component removed; the density wave has a predominantly $d-$ form factor for the values of the parameters chosen. This is qualitatively identical to the density wave observed in at least three different families of the underdoped cuprates \cite{Fujita14,Comin14sym}.

\subsection{Broken rotational symmetry}
When $R_{\pi/2}$ is no longer a symmetry (as would be the case if the parent $\mathbb{Z}_2$ spin liquid state had broken $C_4$ symmetry), the free-energy in the axial case can be modified easily at quadratic order,
\beq
{\cal{F}}_a: \frac{r}{2} \left(\sum_{i=1}^4\psi_i^2\right) \rightarrow \frac{r_x}{2} \left(\psi_1^2+\psi_3^2\right) +\frac{r_y}{2}\left(\psi_2^2+\psi_4^2\right).
\eeq
An obvious consequence of the above modification is that it favors ground states with one ordering direction over the other in phase I, removing the possibility of patterns rotated by $90^\circ$ in Fig. \ref{fig:patterns1}(a),(c) and hence patterns with $(q,0)$ and $(0,q)$ type ordering wavevectors are no longer degenerate. 

In the diagonal case, inversion symmetry demands equivalence under $\{1,3\}\leftrightarrow\{2,4\}$, and hence the quadratic part of the free energy stays the same and the coefficients of some of the quartic terms must be changed instead to break rotation symmetry. The number of quartic couplings in $\mathcal{L}_d$ then increases from 4 to 7. Again, patterns rotated by $90^\circ$ are no longer degenerate, but, unlike the axial case, the degeneracy between patterns with $(q,q)$ and $(q,-q)$ type ordering wavevectors is preserved by inversion symmetry.

\subsection{Monte-Carlo simulations}
\begin{figure*}[!ht]
\includegraphics[height=2.75in]{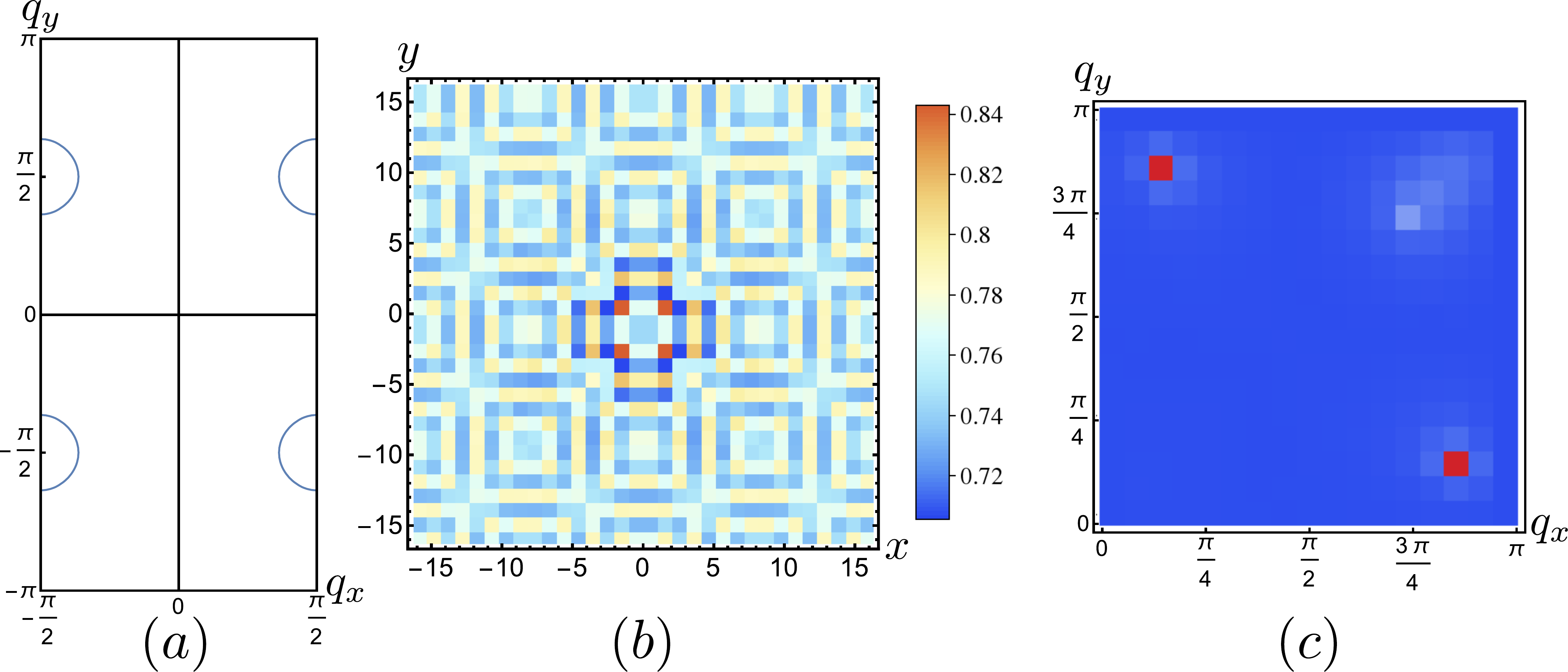}
\caption{(a) The blue arcs denote the soft-spin dispersion minimum for the values of the couplings mentioned in the text. (b) The horizontal-vertical bond energy-energy correlator $\langle E^+_x(0)E^+_y(a)\rangle$ from the Monte Carlo simulation, showing an incommensurate pattern. (c) The magnitude of the Fourier transform of this pattern in the $(+,+)$ momentum quadrant, the other three quadrants are related by rotation.}
\label{fig:mcpatterns}
\end{figure*}

We performed a Monte Carlo simulation of $H=H_0+H_1$ with $J_2=J_3=0$ on a $32\times32$ lattice with periodic boundary conditions. Choosing $J_1\approx1.8,~J\approx3$, the simulation produces weak incommensurate modulation in the energy-energy correlators for the bonds. In the corresponding soft-spin model, the dispersion minima lie on small arcs centered near $(\pm \pi/2, \pm \pi/2)$ (See Fig. \ref{fig:mcpatterns}(a)). However, due to finite size effects, only the wavevectors on these arcs that are nearly commensurate with the square geometry will appear in the ground states. The resulting horizontal-vertical bond energy-energy correlator $\langle E^+_x(0)E^+_y(a)\rangle$  (Fig. \ref{fig:mcpatterns}(b)) shows modulation at multiple incommensurate wavevectors (Fig. \ref{fig:mcpatterns}(c)), which belong to the set of wavevectors $\{\tilde{q}^{(n)}\}$ defined in Eq.~(\ref{eq:rho}) when the wavevectors $q^{(n)}$ of the ground states lie on the arcs. If the values of the $J$'s are reduced to move the minima away from $(\pm \pi/2, \pm \pi/2)$, then the depth of the minima is also simultaneously reduced and lower temperatures would be required to observe incommensurate patterns; however, the single spin flip procedure we used is not ergodic at these lower temperatures due to the energy barriers between degenerate states becoming large relative to the temperature. It would be an interesting subject for future study to map out the phase diagram of this model using more robust Monte Carlo methods, while also including the effects of additional couplings such as $J_2$ and $J_3$.

\section{Conclusion}

Our motivation in this work was to extend the well developed theory of confinement in insulating spin liquids to fractionalized Fermi liquids. One of our main results is that the structure of the transition is remarkably similar to the corresponding transition in the insulator. 
First, the gapless fermions do not lead to Landau damping of the critical modes \cite{Morinari02,Kaul08,TGTS09}.
Second, in the FL* phase, the gapless fermions do not carry a charge under the emergent gauge-field, and hence the PSG transformations for the topological excitations of the underlying gauge-theory remain unmodified. The only allowed couplings between the fermions and the gauge fields
are non-minimal; these lead to long-range and frustrating couplings in the action for the topological excitations, and predict 
a plethora of possibilities for the patterns of broken symmetries in  the confined phase.

Focusing specifically on the $\mathbb{Z}_2$-FL* phase, we studied the patterns of density wave order that arise upon condensing visons. By tuning the relative strengths of the microscopic interactions between the visons, we were able to obtain a unidirectional and incommensurate density wave state with predominantly $d-$form factor upon confinement. A number of experiments have now reported the density wave order in the non-Lanthanum based cuprates to be of this type. It is then natural to ask if the pseudogap metal can be described by a $\mathbb{Z}_2$-FL*, and moreover, if the charge ordering transition in the metallic phase could be associated with a confinement transition of the type studied in this paper. 
Finally, we note that a theoretically challenging task for the future is to describe a transition out of the $\mathbb{Z}_2$-FL* into a metal with a large Fermi-surface, as this may hold the key to understanding the remarkable properties associated with the strange-metal phase.

\section*{Acknowledgments} 

We thank Y. Qi for useful discussions. This research was supported by the NSF under Grant DMR-1360789.
Research at Perimeter Institute is supported by the Government of Canada through Industry Canada 
and by the Province of Ontario through the Ministry of Research and Innovation.

\appendix
\section{Derivation of sine-Gordon theory}
\label{app:duality}
\begin{figure}
\begin{center}
\includegraphics[height=3.0in]{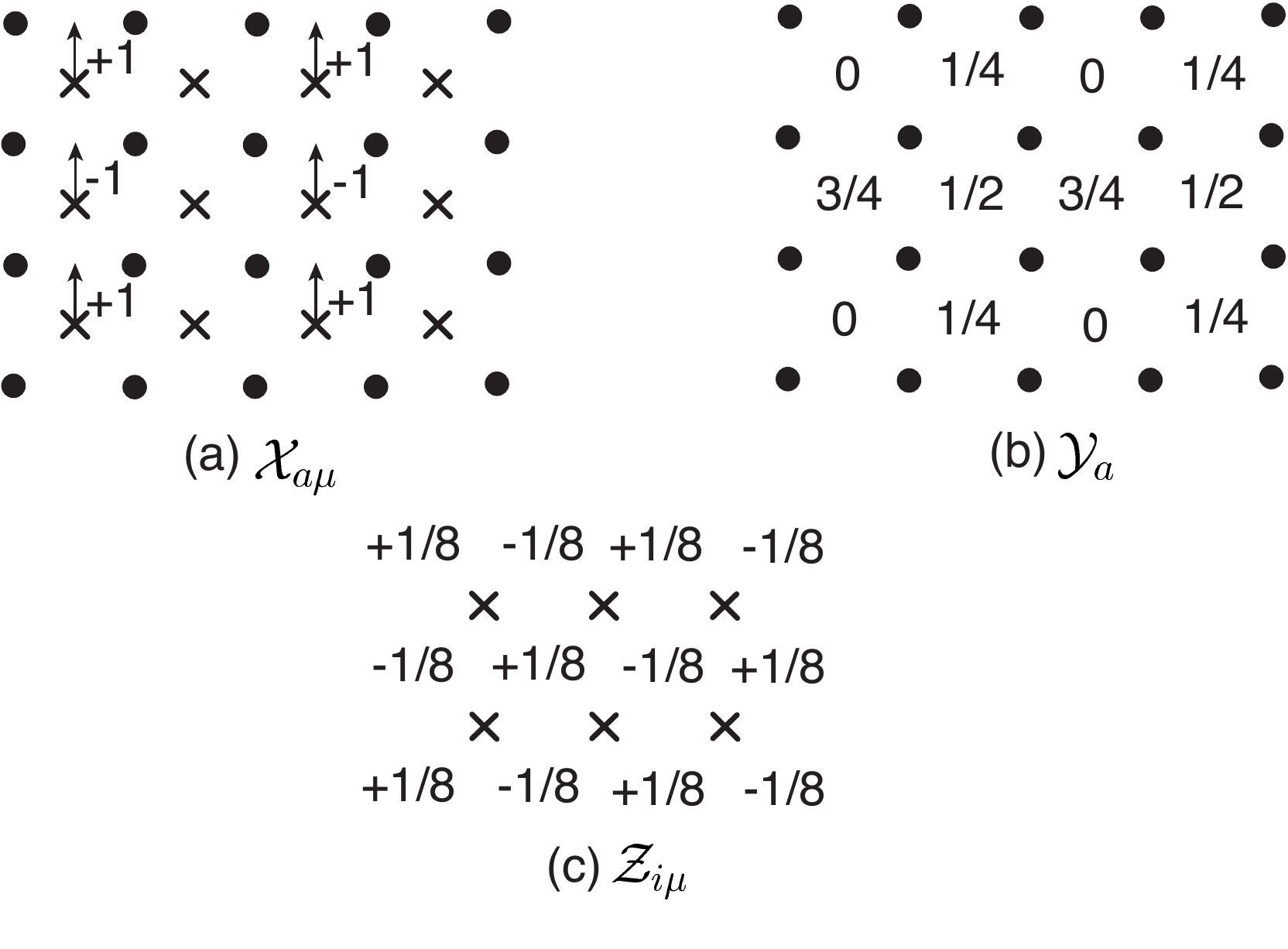}
\end{center}
\caption{The values of the only non-zero components of the fixed
field ${\cal X}_{a \mu}$, ${\cal Y}_a$, and ${\cal Z}_{i \mu}$.
The circles (crosses) are the sites of the direct (dual) lattice.
In (c), only the $\mu=\tau$ component of ${\cal Z}_{i \mu}$ is
non-zero and its values are shown.
}
\label{fig2}
\end{figure}

This appendix is adapted from Ref. \cite{SSMV99} for the case of the fermionic dimer model. We solve for the constraint in Eq.~(\ref{f5}) by writing $F_{a\mu}$ as
the sum of a particular solution and the general solution of the
homogeneous equation:
\begin{equation}
F_{a \mu} = \Delta_{\mu} N_a + 2 S {\cal X}_{a \mu}.
\label{f6}
\end{equation}
Here $N_a$ is a fluctuating integer-valued field on the dual lattice
sites, while ${\cal X}_{a \mu}$ is a fixed field independent of $\tau$
satisfying
\begin{equation}
\epsilon_{\mu\nu\lambda} \Delta_{\nu} {\cal X}_{a \lambda} = \eta_i
\delta_{\mu \tau}.
\label{curlx}
\end{equation}
A convenient
choice is to take ${\cal X}_{a x} = 0$, ${\cal X}_{a \tau} = 0$,
and ${\cal X}_{a y}$
as shown in Fig.~\ref{fig2}(a), taking the values $\pm 1$ on every second column
of sites and zero otherwise.
For future manipulations, it is convenient to split ${\cal X}_{a \mu}$
into curl-free and divergence-free parts by writing
\begin{equation}
{\cal X}_{a \mu} = \Delta_{\mu} {\cal Y}_a + \epsilon_{\mu \nu
\lambda} \Delta_{\nu} {\cal Z}_{i \lambda},
\label{f7}
\end{equation}
where again ${\cal Y}_a$ and ${\cal Z}_{i \mu}$ are fixed fields
independent of $\tau$ and their values are shown in
Fig.~\ref{fig2}(b),(c); ${\cal Y}_a$ takes the values
$0,1/4,1/2,3/4$ on the four dual sublattices, while
${\cal Z}_{i \mu} = \delta_{\mu \tau} \eta_i /8$.
Finally, we promote
the integer valued field $N_a$ to a real-valued field $\varphi_a$
by the Poisson summation formula, and shift the real field
by $\varphi_a \rightarrow \varphi_a - 2 S {\cal Y}_a$. This leads to the theory in Eq.~(\ref{sgf}).

\section{Bond patterns from U(1) FL*}
\label{app:u1p}
We can imagine integrating out the fermions, and obtaining an effective action for the height field on the dual lattice in the form
\beq
H_{\rm eff} &=& \sum_{a_x, a_y} \Bigg[ \frac{K_1}{2} \left( \varphi_{a+2 \hat{e}_x} - \varphi_a \right)^2 + \frac{K_1}{2} \left( \varphi_{a+2 \hat{e}_y} - \varphi_a \right)^2 - y \cos(2 \pi(\varphi_a -2 S \mathcal{Y}_a)) \nonumber \\
&+& \frac{K_2}{2} \left( \varphi_{a+2 \hat{e}_x+2 \hat{e}_y} - \varphi_a \right)^2 + \frac{K_2}{2} \left( \varphi_{a+2 \hat{e}_x - 2 \hat{e}_y} - \varphi_a \right)^2 \nonumber \\
&+& \frac{K_3}{2} \left( \varphi_{a+4 \hat{e}_x} - \varphi_a \right)^2 + \frac{K_3}{2} \left( \varphi_{a+4 \hat{e}_y} - \varphi_a \right)^2 \nonumber  \\
&+& \frac{K_4}{2} \left( \varphi_{a+4 \hat{e}_x+4 \hat{e}_y} - \varphi_a \right)^2 + \frac{K_4}{2} \left( \varphi_{a+4 \hat{e}_x - 4 \hat{e}_y} - \varphi_a \right)^2  \nonumber \\
&+& \frac{K_5}{2} \left( \varphi_{a+2 \hat{e}_x} - \varphi_a \right)^4 + \frac{K_5}{2} \left( \varphi_{a+2 \hat{e}_y} - \varphi_a \right)^4 \Bigg]. 
\label{eq:extdhtmdl}
\eeq

\begin{figure}[!ht]
\includegraphics[height=5.0in]{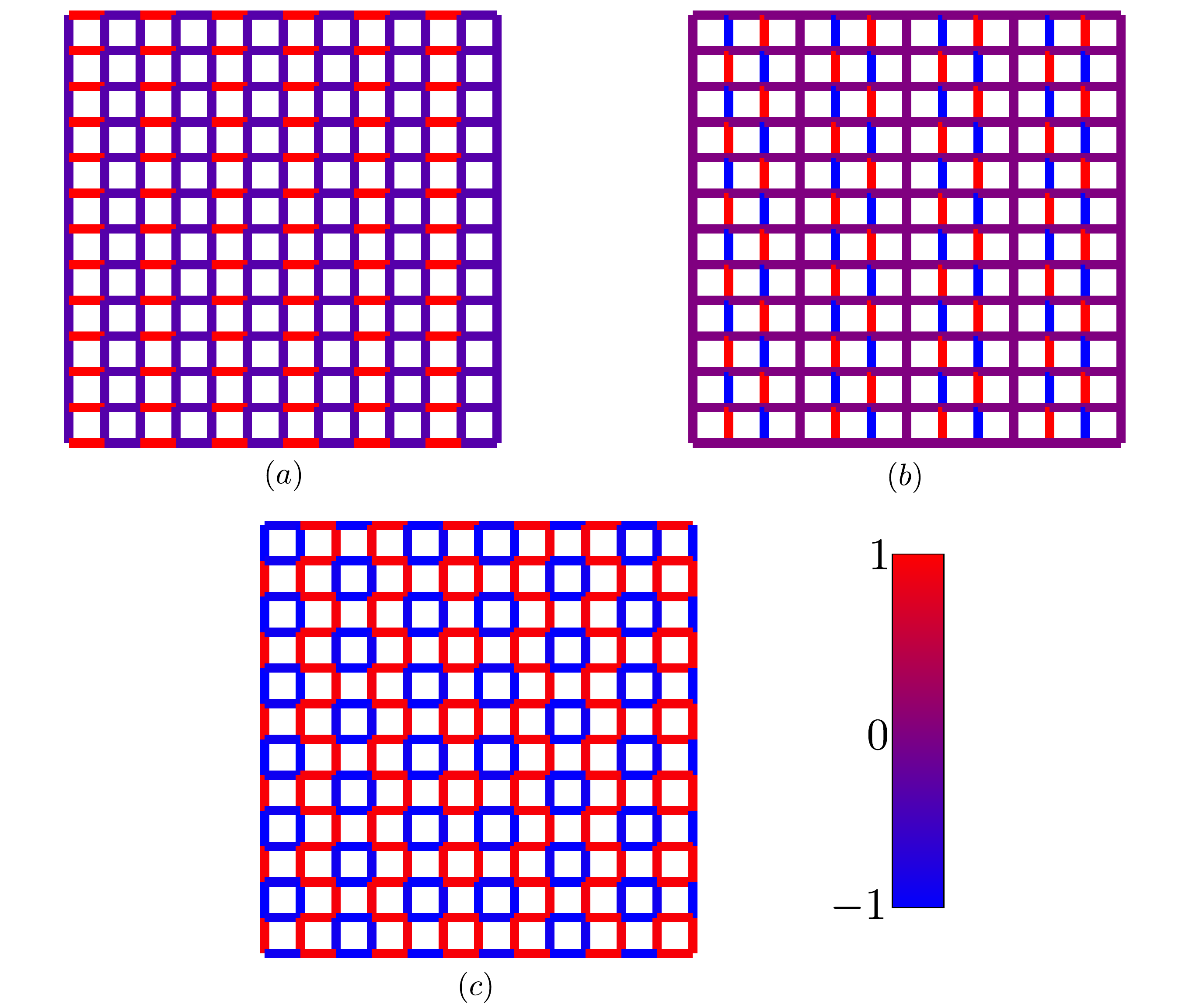}
\caption{Individually normalized bond density wave patterns on the direct lattice for the $U(1)$ version of the problem discussed above. (a) The columnar VBS state that results when $y\neq0$ and the dispersion minimum is at zero momentum. Modulated density waves are produced when the dispersion minimum is not at zero momentum. We used (b) $K_1=0.25,~K_2=-0.5,~K_3=K_4=0,~K_5=0.025,~y=0$ and (c) $K_1=0.25,~K_2=-0.5,~K_3=K_4=0,~K_5=0.025,~y=0.25$.}
\label{fig:U1patterns}
\end{figure}

As in the $\mathbb{Z}_2$ case, the bond observables are determined by the terms in the height field Hamiltonian coupling to the dimer density on the direct lattice bonds;
\beq
\rho_y^+(a) \propto (-1)^{a_x + a_y} \left( \varphi_{a + 2 \hat{e}_x} - \varphi_a \right),~~\rho_x^+(a) \propto -(-1)^{a_x + a_y} \left( \varphi_{a + 2 \hat{e}_y} - \varphi_a \right).
\eeq

We then proceeded to minimize Eq.~(\ref{eq:extdhtmdl}) numerically on a $24\times24$ lattice with periodic boundary conditions. When $K_2,~K_3,~K_4$ are sufficiently negative, the $\varphi$ dispersion has its minima at nonzero momenta, and modulated states are produced. This happens, for example when $K_1 + 2K_2 < 0,~K_3=K_4 = 0.$ (axial wavevectors) or when $K_1+4K_3<0,~K_2=K_4 = 0$ (diagonal wavevectors) or when $K_1+8K_4<0,~K_2=K_3 = 0$ (axial wavevectors). In these cases we need $K_5>0$ to stabilize the free energy. When none of these conditions are true, the dispersion minimum is at zero momentum, and the lowest energy state is the columnar VBS state shown in Fig. \ref{fig:U1patterns}(a) for $y\neq0$.

Fig. \ref{fig:U1patterns}(b) shows the bond pattern produced when $K_1 + 2K_2 < 0,~K_3=K_4 = 0.$ and $y=0$ (deconfined phase). The values of $K_1,~K_2$ are chosen so that the dispersion minima are located at $(0,\pm\pi/3)$ and $(\pm\pi/3,0)$. The state displayed has modulation wavevectors $(\pm\pi/3,0)$ (a degenerate state has modulation wavevectors $(0,\pm\pi/3)$. When $y$ is increased, a complicated bond pattern with additional wavevectors near $(0,\pi)$ and $(\pi,0)$ are produced, with a multitude of form factors (Fig. \ref{fig:U1patterns}(c)). In the limit of $y\rightarrow\infty$, we recover the columnar VBS state of Fig. \ref{fig:U1patterns}(a).

We also considered the possibility of ``tilt" phases, implemented by allowing for boundary conditions of the type
\beq
\varphi_{a+2L\hat{e}_x} = \varphi_a + t_x,~~\varphi_{a+2L\hat{e}_y} = \varphi_a + t_y,
\eeq
where $t_{x,y}$ are integers. When $t_{x,y}=0$ we have periodic boundary conditions. For all cases tested, we found that allowing for $t_{x,y}\neq0$ increases the energy of the ground states, indicating that the ``tilt" phases \cite{AVLBTS04, EFSLS04} are energetically unfavorable.

\section{Eigenmodes}
\label{em}
The eigenmodes corresponding to the vison dispersion eigenvalues (Eq.~(\ref{eq:vdisp})) are
\beq
v^\pm(q_x,q_y)=\frac{|\cos q_y|}{(L_xL_y)^{1/2}\sqrt{\cos^2q_y+(\cos q_x+ \xi^\mp_0(q))^2}}\left(\frac{\cos q_x+ \xi^\mp_0(q)}{|\cos q_y|}e^{i q\cdot a} + \mathrm{sgn}(\cos q_y)e^{i (q+K_x)\cdot a}\right),\nonumber\\
\eeq 
and hence their PSG transformation rules are
\begin{align}
&T_x : v^\pm(q_x,q_y) \rightarrow e^{-iq_x}v^\pm(q_x,q_y+\pi),~T_y :  v^\pm(q_x,q_y) \rightarrow e^{-iq_y}v^\pm(q_x,q_y), \nonumber \\
&I_x :  v^\pm(q_x,q_y) \rightarrow v^\pm(q_x,-q_y),~I_y : v^\pm(q_x,q_y) \rightarrow v^\pm(-q_x,q_y), \nonumber \\
&R_{\pi/2} : v^\pm(q_x,q_y) \rightarrow \mp \frac{v^\pm(-q_y, q_x) + v^+(-q_y,\pi+q_x)}{\sqrt{2}},~|q_y|<\pi/2, \nonumber \\
&R_{\pi/2} : v^\pm(q_x,q_y) \rightarrow \mp \frac{v^\pm(-q_y+\pi, q_x) - v^\pm(-q_y+\pi,\pi+q_x)}{\sqrt{2}},~|q_y|>\pi/2.
\label{eq:psgv}
\end{align}

\bibliographystyle{apsrev4-1_custom}
\bibliography{refs}
\end{document}